\documentclass{article} 
\usepackage{colm2024_conference}

\usepackage{microtype}
\usepackage{hyperref}
\usepackage{url}
\usepackage{booktabs}
\usepackage{tikz}
\usepackage{xcolor}
\usepackage{mdframed}
\usepackage{amssymb}
\usepackage{rotating}
\usepackage{booktabs}
\usepackage{array}

\title{Language Models as Critical Thinking Tools:\\ A Case Study of Philosophers}

\author{Andre Ye$^{\alpha\gamma}$, Jared Moore$^\beta$, Rose Novick$^\gamma$, Amy X. Zhang$^\alpha$ \\
\small \texttt{andreye@uw.edu}, \texttt{jlcmoore@stanford.edu}, \texttt{amnovick@uw.edu}, \texttt{axz@cs.washington.edu} \\
\small $^\alpha$ Paul G. Allen School of Computer Science and Engineering, University of Washington \\
\small $^\beta$ Department of Computer Science, Stanford University \\
\small $^\gamma$ Department of Philosophy, University of Washington
}

\colmfinalcopy 
\begin{document}

\maketitle

\begin{abstract}
Current work in language models (LMs) helps us speed up or even skip thinking by accelerating and automating cognitive work.
But can LMs help us with \textit{critical thinking} --- thinking in deeper, more reflective ways which challenge assumptions, clarify ideas, and engineer new concepts?
We treat philosophy as a case study in critical thinking, and interview 21 professional philosophers about how they engage in critical thinking and on their experiences with LMs.
We find that philosophers do not find LMs to be useful because they lack a sense of selfhood (memory, beliefs, consistency) and initiative (curiosity, proactivity). We propose the \textit{selfhood-initiative} model for critical thinking tools to characterize this gap.
Using the model, we formulate three roles LMs could play as critical thinking tools: the Interlocutor, the Monitor, and the Respondent.
We hope that our work inspires LM researchers to further develop LMs as critical thinking tools and philosophers, and other `critical thinkers' to imagine intellectually substantive uses of LMs.
\end{abstract}

\newcommand{\andre}[1]{\textcolor{blue}{[andre: #1]}}
\newcommand{\jared}[1]{\textcolor{blue}{[jared: #1]}}
\newcommand{\rose}[1]{\textcolor{blue}{[rose: #1]}}
\newcommand{\amy}[1]{\textcolor{blue}{[amy: #1]}}

\newcommand{\name}[1]{\textcolor{red}{[#1]}}
\newcommand{\bluecircle}[1]{%
    \tikz[baseline=(char.base)]{
        \node[shape=circle, draw, inner sep=0.5pt, 
            fill=blue!20, text=black, font=\sffamily\bfseries] (char) {#1};
    }%
}
\newcommand{\redcircle}[1]{%
    \tikz[baseline=(char.base)]{
        \node[shape=circle, draw, inner sep=0.5pt, 
            fill=red!20, text=black, font=\sffamily\bfseries] (char) {#1};
    }%
}
\newcommand{\icite}[1]{%
    \textcolor{gray}{(#1)}%
} 
\newcommand{\g}[1]{%
    \textcolor{gray}{#1}%
} 

\newmdenv[leftline=true, linecolor=black, topline=false, bottomline=false, rightline=false]{leftbarquote}



\section{Introduction}

\begin{leftbarquote}
\textbf{\textit{``But I like the inconveniences.''}} ---
``We don't,'' responds the Controller. ``We prefer to do things comfortably.'' --- ``But I don’t want comfort,'' John gasps. ``I want God, I want poetry, I want real danger, I want freedom, I want goodness. I want sin.'' --- ``In fact,'' says the Controller, ``you’re claiming the right to be unhappy ... the right to live in constant apprehension of what may happen tomorrow; ... the right to be tortured by unspeakable pains of every kind.''
There is a long silence.
``I claim them all,'' says John at last.~ (\textit{Minimally adapted from \cite{huxley1932brave}}.)
\end{leftbarquote}

Language Models (LMs) have recently alleviated a whole host of our intellectual inconveniences.
They can help us do things we would have begrudgingly done by ourselves otherwise: write code~\citep{chen2021evaluating,rozière2024code}, generate emails ~\citep{Goodman2022LaMPost}, and translate text ~\citep{nllbteam2022language}.
In sparking ideas by generating stories~\citep{schwitzgebel2023creating} and concept designs~\citep{10.1145/3582269.3615596}, LMs offer shortcuts to gaining new thoughts. 
They also help us put our thinking into words by revising~\citep{Mysore2023PEARLPL} and giving feedback~\citep{doi:10.1056/AIoa2400196} on our writing.
In all these cases, LMs help us speed up and circumvent the inconveniences of thinking ourselves.

In many contexts, however, the ``inconvenience'' of thinking is not a temporary problem to be alleviated, but a deep puzzle to be reflected upon.
Many people are invested in specific areas of intellectual inquiry --- e.g., historians, scientists, philosophers --- and more generally, in reflection and engagement with the world --- e.g., as informed political citizens, critical information consumers, and moral actors.
They are interested in identifying and challenging assumptions, clarifying muddled ideas, and engineering new and useful ways to think.
Core to this sort of inquiry is \textit{critical thinking} --- ``the propensity and skill to engage in an activity with reflective skepticism''~\citep{mcpeck2016critical}.
Can LMs serve as tools for \textit{critical thinking} --- helping us think more deeply and in more complex ways, rather than faster or not at all?
What if --- like John --- \textit{we claim all the rights to think}~\citep{10.1145/3449287}?

To investigate how LMs can serve as critical thinking tools, we use philosophers as a case study --- philosophers being people who are in the business of thinking critically about a wide range of concepts and ideas.
We interview \textbf{21 professional philosophers} to understand their thinking processes, collect their experiences with and views on current LMs, and brainstorm the roles LMs could play as critical thinking tools in philosophy (\S\ref{methods}).
We find that current philosophers \textit{do not} think LMs are good critical thinking tools (\S\ref{bad-critical-thinking}) for two primary reasons: LMs are too neutral, detached, and nonjudgmental (\S\ref{view-nowhere}); and LMs are too servile, passive, and incurious (\S\ref{thought-mirror}).
We propose the \textit{selfhood-initiative} model for critical thinking tools, which explains why philosophers find conversations with other philosophers and reading philosophical texts to be more helpful for their work than current LMs~(\S\ref{selfhood-initiative}).
Using the model, we describe \textbf{three roles} LMs could play as critical thinking tools: the Interlocutor, the Monitor, and the Respondent~(\S\ref{roles}).
Finally, we outline how these LMs could inform metaphilosophical questions and shape the discipline of philosophy~(\S\ref{metaphilosophy}), and discuss challenges in building LMs~(\S\ref{lm-challenges}) and interfaces~(\S\ref{hai-challenges}) for critical thinking.


\section{Background and Related Work}

\subsection{Language Models as Thinking Tools}

A large and growing literature investigates how LMs can serve as thinking tools for humans engaged in intellectual work.
This research tends to concern how LMs can serve two intellectual functions: \textit{idea stimulation} (roughly, ``divergent thinking'') and \textit{idea refinement} (roughly, ``convergent thinking'')~\citep{banathy1996designing, designcouncil_double_diamond_2019}.

LMs can provide \textit{stimulus for ideas} --- information and (re)formulations which provoke and guide creative processes.
Generally, LMs can expand idea sets~\citep{DiFede2022TheIM}, produce creative analogies~\citep{Bhavya2023CAM} and metaphors~\citep{Chakrabarty2021MERMAIDMG}; discover concepts~\citep{Lam2024LLooM}, and facilitate group ideation~\citep{Rayan2024, shaer2024aiaugmentedbrainwritinginvestigatinguse}.
LMs to open-endedly propose plots, characters, and entire stories for creative writers~\citep{Calderwood2020HowNU, Schmitt2021CharacterChatST, 10.1145/3490099.3511105, Mirowski2022CoWritingSA, 10.1145/3491102.3501819, Chakrabarty2023CreativitySI}; but also provide inspiration in more constrained creativity tasks, such as science writing~\citep{Gero2021SparksIF, kim2023metaphorian}.
Although fraught with pitfalls~\citep{Messeri2024}, scientists can use LMs to find and synthesize literature~\citep{van2021automation, wagner2022artificial, Fok2023,  Khraisha2024} and iterate through research inquiry~\citep{Wang2023, morris2023scientists, liu2024ai}.
Designers can use LMs to generate and develop concept designs 
~\citep{10.1145/3582269.3615596, 10.1145/3563657.3596098, ma2023conceptual, chong2024cadpromptedgenerativemodelspathway, ma2024didupdynamiciterativedevelopment, Chen2024}.


On the other hand, LMs can also aid the \textit{refinement of ideas} -- selecting from and improving upon an established existing pool of ideas.
LMs can help writers by making suggested revisions~\citep{Du_2022, Zhao2022LeveragingAI, Mysore2023PEARLPL, shu2023rewritelm} and clarifying writing goals~\citep{Arnold2021GenerativeMC, kim2024authorship}.
For scientists, LMs can facilitate revision of scientific writing~\citep{doi:10.1056/AIoa2400196, radensky2024letspointllmsupportedplanning}; for designers, LMs can provide feedback on~\citep{10.1145/3613904.3642782} and annotate~\citep{lu2024flowysupportinguxdesign} UIs.
In teaching settings, writing feedback given by LMs may be more motivating~\citep{MEYER2024100199} and engaging~\citep{tanwar2024opinebot} than feedback given by other humans.
Besides reviewing ideas, LMs can also curate them --- for instance, by summarizing writing~\citep{10.1162/tacl_a_00373, 10.1145/3526113.3545672} and identifying important ideas~\citep{10.1145/3613904.3642217}.


\subsection{Language Models as Critical Thinking Tools}

However, one part of the thinking process is clearly missing.
One does not simply go from the stimulus for ideas to figuring out how to refine them: one needs to do the actual \textit{critical thinking}, involving reflection upon ideas, judgment, and conceptual engineering.
LMs can help provide the seeds for our ideas when we don't have any (i.e., stimulus) and help us work through them once we've got them (i.e., refinement), but how can they help us with questioning, reorienting, analyzing, and developing ideas (i.e., critical thinking)?

There are many different definitions of critical thinking:
``the propensity and skill to engage in an activity with reflective skepticism''~\citep{mcpeck2016critical}, 
``reasonable, reflective thinking that is focused on deciding what to believe or do''~\citep{ennis1993critical}, and
``the development and evaluation of arguments''~\citep{facione1984toward}, among many others.
Critical thinking requires many dispositions, such as seeking clear statements of questions, looking for alternatives, and being open-minded~\citep{ennis1987taxonomy}.
Critical thinking is what makes many areas of intellectual inquiry --- such as history, science, and philosophy --- difficult.
In these areas, people must produce and work with observations that are incomplete and open to a multiplicity of framings to pursue problems with often unclear definitions of progress --- a landscape which demands critical thinking.
For instance, on different accounts, history requires interpreting the past with alternative (nonlinear, long-range) temporalities~\citep{Lagrand2016TheMA}, taking into account the ways in which power structures shape historical record and memory~\citep{Foucault1995TheAO, Winichakul1997MichelRolphTS}, and identifying and manipulating narrative structures~\citep{White1975MetahistoryTH,Pauly2005JohnLG}. Science requires advances not only in empirical work, but also reflection upon underlying paradigms of research~\citep{Kuhn1963TheSO}, epistemology~\citep{harding2013rethinking}, and the social and material factors that constitute scientific knowledge~\citep{Latour1989ScienceIA}.

Researchers across a variety of fields have developed a rich tapestry of approaches and tools to support critical thinking and related acts.
Educators develop teaching strategies to promote critical thinking~\citep{mcpeck1990teaching, pithers2000critical} such as teaching and interlinking a variety of perspectives on a subject in an integrative manner~\citep{enciso2017critical} and encouraging students' intellectual independence in finding answers to their questions~\citep{langer1997power, raths1966teaching}.
Psychologists and cognitive scientists seek to understand how cognitive mechanisms and biases inform how humans (should) develop critical thinking~\citep{carey1986cognitive, reif2008applying}, emphasizing the cultivation of basic metacognitive ``building blocks'' of critical thinking~\citep{pasquinelli2021naturalizing} and
teaching for ``practical theory''~\citep{gelder2005teaching}.
Meanwhile, human-computer interaction (HCI) researchers explore how interactions with computer applications can facilitate critical thinking:
designers can provoke experiences of discomfort~\citep{Benford2012, 10.1145/2675133.2675162}; 
emphase understanding over rote expression in social contexts~\citep{10.1145/2207676.2208621, 10.1145/3027063.3053250}; 
and build small ``nudges''~\citep{10.1145/3491101.3519815} into interfaces which ``prime''~\citep{10.1145/3176349.3176377} users towards reflective critical thinking~\citep{Bentvelzen2022};
among many others.
Many of these themes will be revisited in our discussion of design proposals for LMs as critical thinking tools~(\S\ref{roles}).

A growing body of work has explored how LMs might contribute towards critical thinking.
LM-based news and media can positively affect users' willingness to think through opposing or novel viewpoints, which can be applied to combat polarization and extremism~\citep{10.1145/3613904.3642513, doi:10.1177/0267323120940908, doi:10.1177/1461444821993801, Wang2022, Blasiak2021}.
\cite{cai2024antagonistic} consider how currently ``sycophantic'', ``servile'', and ``lobotimized'' LMs can be used in more critical ways by challenging users' pre-existing ideas and constructively using antagonistic interactions to develop their thinking.
\cite{Danry2023, ma2023conceptual,  park2024thinking} show how LMs can facilitate human self-reflection and improve human reasoning by asking questions instead of only answering them (as in the typical LM interaction paradigm).
\cite{10.1145/3640543.3645196} encourage critical thinking by building LM interactions using structured templates (over free-form chat).
In more targeted contexts,
LMs can be used to help scientific researchers critically think about their impact statements~\citep{impactbot}, and to help political theorists to metacognitively reflect upon their own creative processes and judgments~\citep{Rodman2023}.





\subsection{Philosophy as Critical Thinking, Critical Thinking as Philosophy}

In this paper, we focus on philosophy as a case study for critical thinking.
Philosophy is concerned with critical, systematic, and reflective examination of the world.
This includes understanding the basic structure of life and the world --- what does it mean to exist~\citep{Aristotle1908Metaphysics,Heidegger1962BeingTime,Sartre1993Being}, live~\citep{Aurelius2006Meditations}, and die~\citep{Kierkegaard1983Sickness,Nietzsche1961Zarathustra}?; 
what does it mean to know something~\citep{Plato2008Theaetetus, kant1781critique, HusserlCairnsPhenomenology} and what are the limits of scientific knowledge ~\citep{Popper2002Logic,Chalmers2013Science}?;
on what moral bases should we act~\citep{Aristotle2004Nicomachean,Spinoza2003Ethics}, and is it even possible to determine `objective' answers to moral questions~\citep{Hume2003Treatise,Harman1996Moral, foucault1990use}?
Core to philosophy is ``the endeavour to know how and to what extent it might be possible to think differently, instead of legitimating what is already known''~\citep{foucault1990use}.
Philosophy is for intellectual creation and engineering: \cite{Deleuze1991} wrote that ``So long as there is a place for creating concepts, the operation that undertakes this will always be called philosophy.''
In thinking about how to think, philosophy is not only about \textit{suspicion} toward the meanings and functions of phenomena, but also \textit{recovery} of new significances and coherence~\citep{ricoeur1981hermeneutics}.

Contrary to the image that philosophy is ``done in the armchair'', isolated and impractical, philosophy has always been intertwined with other lines of inquiry.
Plato engaged extensively with advanced mathematics; Aristotle contributed to early physics; Hume leaned on psychology.
Philosophy has asked and continues to ask urgent, relevant questions:
for instance,
how are we to understand the strangeness of quantum mechanics in physics~\citep{Carnap1966}; the relationship between consciousness (mind) and the brain (matter)~\citep{Chalmers2013Science}; and ``fairness'' and ``justice'' in contexts like algorithmic discrimination~\citep{Hu2023-HUWIRJ}, legal punishment~\citep{Alexander1922Philosophy}, and the distribution of resources~\citep{41832fbd-4a82-3c18-ae99-e02c0f77c64f}?
Indeed, researchers in every area of intellectual inquiry confront philosophical questions in their work: 
they might ask if a model or concept is ``really real'' (and how they know so),
what the ``nature'' of their object of study is,
aim to formulate normative desiderata for their theories,
and so on.
Therefore, we study philosophers' views and practices in this paper both because philosophers engage extensively in critical thinking \textit{and} because many questions which require critical thinking asked by non-philosophers often have a philosophical flavor.

\section{Methods}
\label{methods}

The first author conducted interviews with 21 professional philosophers at 14 philosophy departments at doctoral universities in the United States.
We contacted and selected philosophers for high diversity across area of interest (e.g., ethics, political philosophy, philosophy of science).
Interviews took place online and lasted between 30 to 60 minutes, depending on interviewee availability.
Interviewees were asked how they philosophize (e.g., where ideas come from, how ideas are developed, what resources are needed) and their views on LMs (e.g., can LMs `do' philosophy, how might they be useful for philosophizing).
These questions followed a loose script (see \S\ref{interview}), although we asked novel follow-up questions to pursue interesting lines of inquiry raised by the interviewees' responses.
In cases where interviewees had very little or no prior exposure to LMs, they interacted live with the GPT-4 model on a philosophical topic of their choosing.
We received IRB approval from our university to conduct the interviews; all interviewees confirmed their consent to participate in the study, and for their responses to inform the development of this paper.
We qualitatively analyzed interview recordings and transcripts.
Using an inductive approach~\citep{Thomas2006} and open coding~\citep{Charmaz2006ConstructingGT}, we identified common themes and positions (yielding \S\ref{bad-critical-thinking} and \S\ref{designing-critical-thinking}).
We refer to interviewees with a unique identifier, e.g., \icite{P1, P2, P3} (see \S\ref{info-sheet}).


\section{Language Models Are Not Good Critical Thinking Tools (So Far)}
\label{bad-critical-thinking}

Many of the interviewed philosophers find LMs to be relevant and interesting, and some find them to have limited uses such as for undergraduate instruction~\icite{P1, P13, P20} or becoming acquainted with a topic~\icite{P5, P11, P12}.
However, none of the philosophers were convinced that current LMs can reliably and conveniently assist them in the intellectually substantive ways which require critical thinking.
Philosophers described current LMs as ``boring'' \icite{P2}, ``anodyne'' \icite{P4}, ``bland'' \icite{P9}, and ``cowardly'' \icite{P13}.
We discovered two broad reasons for this.
First, current LMs tend to be highly neutral, detached, and non-judgmental, often commenting on ideas in abstract and decontextualized ways~(\S\ref{view-nowhere}).
Second, current LMs tend to be servile, passive, and incurious, which is unhelpful when the user does not yet have a clear vision of what they want to accomplish, restricting the variety of intellectual interactions possible
~S\ref{thought-mirror}).

\subsection{How do philosophers philosophize?}
\label{how-do-philosophers}

A close investigation of how philosophers think through difficult philosophical questions can give us insight into the types of tools and interactions which support difficult critical thinking, and provide contrast with current LMs, which fail to perform the same function.

\textit{Where do philosophical ideas come from?}
Philosophers report that their ideas usually come from observing puzzles and tensions in the world, in which some aspect feels bothersome~\icite{P5, P12, P20}, incomplete~\icite{P10, P14}, in need of clarity~\icite{P1, P13}, or outright incorrect~\icite{P3}.
Philosophers encounter these puzzles and tensions most commonly in open conversation with others~\icite{P1, P2, P5, P9, P19} and while reading texts --- books, papers, and monographs making explicitly philosophical arguments or touching upon philosophical themes~\icite{P4, P7, P10, P12, P13, P20}.
These puzzles may have an intellectual or logical character: terms might not be sufficiently disambiguated, inferences may not be valid, and propositions may entail absurd conclusions~\icite{P8, P11}.
However, for many, these tensions are identified and drawn out by ethical motivations~\icite{P1, P8, P16, P12}.
Tensions might arise not primarily because a proposition is incoherent, but rather because it appears ethically problematic. For instance, the trolley problem dilemma was used to probe the differences between doing and allowing harm, with applications to bioethics, particularly abortion~\citep{Foot1967-FOOTPO-2}.
Several philosophers describe being inspired by texts communicating empirical work, seeking to provide explanations for empirical observations~\icite{P1, P2, P16, P18} as well as subjecting the practices and products of the empirical sciences to critical inquiry~\icite{P2, P7, P12, P13, P18}.

\textit{What do philosophers want out of their ideas?}
Once philosophers identify puzzles from conversations and texts, they aim to develop ideas which make progress on these puzzles.
Progress is conceived of in many ways: \textit{``understand[ing] some part of the world better''} \icite{P3}, working through new ways to think about problems \icite{P17}, and better understanding the current ways we think --- for instance, by making implicit assumptions explicit and recognizing the implications of propositions~\icite{P7}.
Some philosophers describe a developed philosophical idea as a ``picture'' \icite{P9, P10} which organizes subideas in a systematic way, allowing one to clearly see the main point(s).
This often requires ``conceptual engineering''~\icite{P6}: challenging, disassembling, and rebuilding the ways in which we think.

\textit{The role of texts in philosophical development.}
Texts continue to actively support the philosophical development past the inception of the idea.
Revisiting texts with an idea in mind can unearth new aspects of the text which comment on that idea~\icite{P9}, and repeatedly consulting written ideas can be helpful for putting words to newly developed ideas~\icite{P2, P20}.
Because texts are static and highly accessible by many people, texts can become a shared basis for and markers in conversation with others~\icite{P9, P19}.
Moreover, because published texts are usually produced by people who have given a problem substantial time and thought, philosophers might approach them with more trust and charity~\icite{P4}.



\textit{The role of conversation in philosophical development.}
Conversations with fellow philosophers are central to evaluating the coherence of ideas~\icite{P21}, raising connections to other ideas and problems~\icite{P5}, and collecting feedback~\icite{P3, P10}.
Conversations may force philosophers to explain and justify ideas they may have taken for granted~\icite{P1}.
Conversation helps philosophers gain confidence that their ideas are good intellectual contributions~\icite{P2, P21}.
Philosophers even simulate conversations in their head, taking on various positions for and against their ideas~\icite{P1, P12}.
Good philosophical conversation requires several conditions.
The interlocutor should be charitable --- genuinely listening to and working through ideas~\icite{P1, P12}, and trusting~\icite{P6, P14} --- but also willing to boldly push ideas forward~\icite{P3} and take intellectual risks~\icite{P18}.
Conversations may not be directed towards any clear goal; interlocutors must be able to ``\textit{riff off each other}''~\icite{P8} and be willing to operate without a preset agenda~\icite{P3, P4}.
This requires interlocutors to be curious about addressing problems~\icite{P21}; it should be a collaborative effort, rather than a combative debate~\icite{P3, P7}.

\subsection{Language Models are neutral, detached, and nonjudgmental}
\label{view-nowhere}

Philosophers find intellectual value when the conversations and texts they encounter provide substantive and well-defended perspectives, but find that LMs do not do the same.

\bluecircle{1} \textit{LMs are abstract, imprecise, and `skirt by' questions.}
Because philosophy is interested in clearly stating and reflecting upon ideas, philosophers often place high value on precision in language.
Changes to a formulation which seem trivial to a layperson may introduce important shifts in meaning for a philosopher.
Meanwhile, LMs seem as if they `tell the user what they want to hear', resulting in risk-averse and hand-waving behavior which produced abstract, imprecise, and ultimately intellectually uninteresting statements~\icite{P5, P7, P15}.
Interviewees noted that when they brought up problems with LMs' responses, LMs skirted around the issue, producing superficially convincing corrections without really addressing the provided issue~\icite{P1, P20}.
LMs are highly factually knowledgeable~\icite{P1} but fail to precisely express philosophical ideas; thus, LMs end up reinforcing the status quo rather than proposing substantive and interesting challenges~\icite{P9}. 

\bluecircle{2} \textit{LM responses change too easily and don't have `weight'.}
Several philosophers describe how easy it is for them to talk LMs into contradictions and incoherent outputs in the same session~\icite{P4, P9}.
 LMs make ``kneejerk reactions'' to user concerns and are excellent at effusively apologizing, but don't ``\textit{fully appreciate}'' their mistakes and the user's comments~\icite{P14}.
Moreover, LM responses seem highly sensitive to trivial changes in the prompt, making some philosophers wary of using them at all~\icite{P21}.
The ease with which one can manipulate an LM's output seems to reduce their trustworthiness and value as tools~\icite{P15}.

\bluecircle{3} \textit{LM outputs don't provide judgments.}
LMs often refrain from formulating serious judgments; they try to remain neutral and `see all sides', but end up presenting all sides in placid and uninteresting ways~\icite{P12, P17}.
They tend to refrain from discussing controversial issues~\icite{P4}, which is unfortunate given that philosophy prides itself on clearly thinking about otherwise-taboo topics of controversy.
As such, LMs are perceived as ``\textit{cowardly}'', refusing to take solid positions and, in some sense, echoing the user~\icite{P13}.
\textit{``It [conversations with LMs] ends up being unproductive and unsatisfying... they don't feel like persons because their language is often so bland and impersonal, non-Socratic, generic... they're boring''} \icite{P9}.

\bluecircle{4} \textit{LMs don't have memory and context.}
Shared context from previous interactions with other humans serve to provide context for and situate ideas in conversation, allowing for efficiency of exploration (as already-exhausted ideas are not brought up again)~\icite{P1, P14}.
Because current popular LM interfaces `lose their memory' of previous interactions in different sessions, LMs often produce general and decontextualized responses to user prompts~\icite{P15}.



\subsection{Language Models are servile, passive, and incurious}
\label{thought-mirror}

Philosophers find intellectual value when fellow philosophers develop their own lines of inquiry in conversation and texts, but find that LMs do not do the same.

\redcircle{1} \textit{LMs fail to be useful in open, undetermined contexts.}
LMs enthusiastically make ``\textit{my problem its problem}''~\icite{P11}, but often philosophers do not have their `problem' entirely clearly thought or formulated~\icite{P5}.
For certain basic tasks, ``\textit{`you have certain success metrics in mind, so you go to [an LM]; but what about truly open-ended conversations where you don’t have success conditions already laid out?''}~\icite{P7}
LM answers often feel like they've been `packaged' or return a `processed end result', whereas \textit{``in the doing of philosophy, we want to be open, in service of a larger dialogue --- philosophy as a process rather than as an end product''}~\icite{P5}.
LMs don't seem to have a drive to know the truth or care about convincing people~\icite{P2, P21} --- features which interviewees note energize interactions even when there is no clearly desired product.


\redcircle{2} \textit{LMs restrict the variety of intellectual interaction.}
The ``incuriosity'' of LMs severely limits possible intellectual interactions philosophers can have with it~\icite{P7}.
\textit{``It's a question-answer platform. It won’t follow up with a ``what do you think?'' ``I'm a little puzzled, how it could be?'' ``Oh gosh, how does it work?'' You can’t have a conversation with [an LM] except one which is like an interview.''}
Several philosophers imagine alternative useful LM interactions in which LMs take on more intellectual risks and independent behaviors:
instead of only answering questions, LMs could also ask them~\icite{P12, P17}, or
LMs might behave with hostility and antagonism towards users' ideas~\icite{P6, P8, P11}.


\section{Designing Language Models for Critical Thinking}
\label{designing-critical-thinking}

Thus far, we've introduced the problem of critical thinking and described how current LMs fail to be good critical thinking tools for philosophers.
Here, we set out a formal model to characterize and compare critical thinking tools~(\S\ref{selfhood-initiative}).
This allows us to imagine new roles for LMs, inspired by what makes people and texts useful as critical thinking tools~(\S\ref{roles}).

\subsection{The Selfhood-Initiative Model}
\label{selfhood-initiative}
We use the two broad reasons why LMs fail to be good critical thinking tools in \S\ref{bad-critical-thinking} as the basis for the model's two axes:
current LMs have low \textit{selfhood}, as they are neutral, detached, and nonjudgmental; they have low \textit{initiative}, as they are servile, passive, and incurious.
In particular,
\textit{selfhood} is a resource's ability to have certain locally persistent internal states (such as perspectives, beliefs, opinions, memory) and to consistently use them as the basis for judgments.
The resource's internal states may change over time due to new knowledge and experiences, but in an intentional and logical (rather than an arbitrary and capricious) manner. Current LMs exhibit \textit{low selfhood} (\S\ref{view-nowhere}).
\textit{Initiative} is a resource's ability to set its own intentions and goals, possibly different from its user's, and to execute actions oriented towards those intentions.
High-initiative resources are not strictly or existentially bound to their user's directives, and may deviate from them.
Current LMs exhibit \textit{low initiative} (\S\ref{thought-mirror}).
These two axes form the \textit{selfhood-initiative} model for critical thinking tools.
Our model is distinct from previous models proposed for the study of critical thinking in that (a) we model \textit{types of critical thinking tools} rather than the \textit{(human) process of critical thinking}~\citep[\textit{inter alia}]{Schon1987, shneiderman2000creating}, and (b) we explore the \textit{interaction} between selfhood and initiative, which have each independently been explored in some capacity by others~\citep[\textit{inter alia}]{cai2024antagonistic, guo2024largelanguagemodelbased, serapiogarcía2023personalitytraitslargelanguage}.
Our model explains why philosophers find texts and other people (but not LMs) to be useful tools, and further provides a design space for LMs as critical thinking tools~(\S\ref{roles}).

\begin{figure}[!t]
    \centering
    \includegraphics[width=\textwidth]{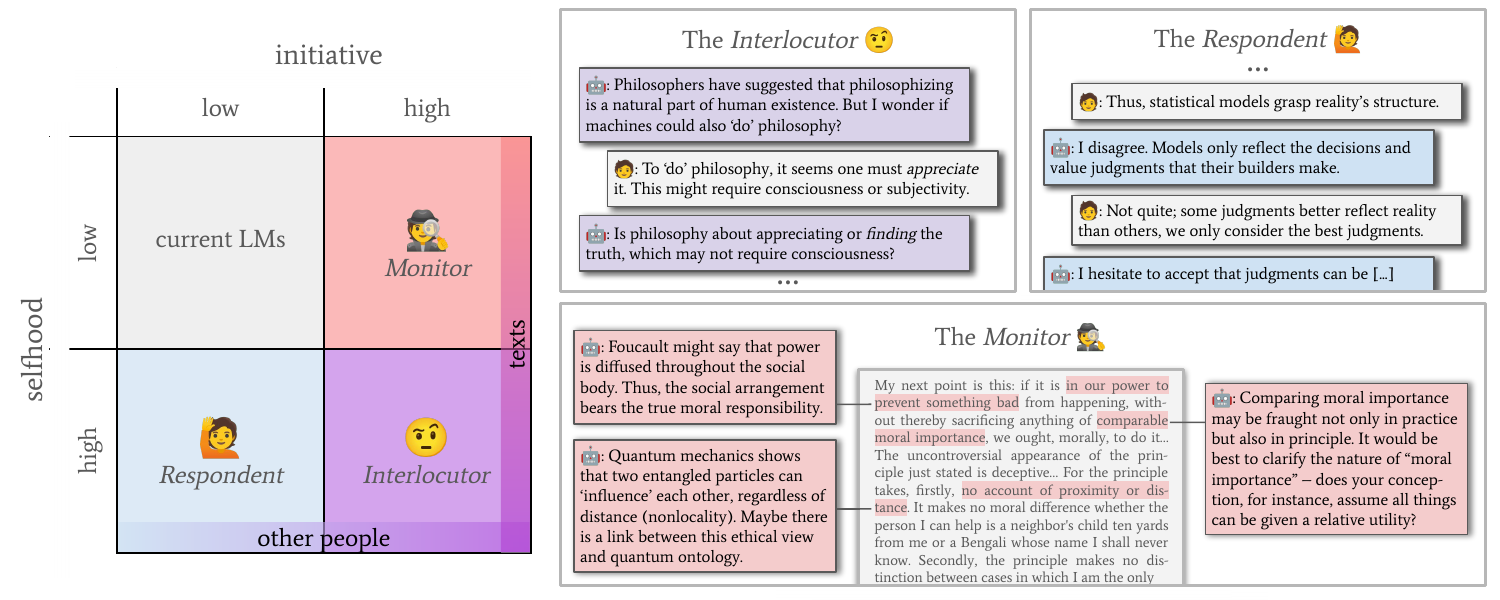}
    \caption{\textit{Left} -- The selfhood-initiative model for critical thinking tools. \textit{Right} -- Illustrative sample interactions between humans and LMs playing different roles. Other alternatives are possible. The excerpt from ``The Monitor'' is taken from \cite{68204752-8786-318f-a155-4726d074c535}.}
    \label{model-roles}
\end{figure}

\textit{Why do philosophers find other people and texts to be useful critical thinking tools?}
In the selfhood-initiative model, other people are \textit{high-selfhood, variable-initiative tools}.
People have specific backgrounds and experiences which inform their views, perspectives, and beliefs; these influence how they understand and respond.
Philosophers find value in talking to other people often \textit{because} of their selfhood; they expect that they will receive interesting judgements and comments, rather than placid neutrality.
However, these people may have variable initiative, depending on the situation.
In free-flowing conversation, each interlocutor may carry the conversation in some direction, whereas in a more focused conversation aimed at collecting feedback, an interlocutor may be expected to directly respond to one's ideas and requests without their own intellectual initiative.
The high selfhood of other people is helpful because it provides particular perspectives and ways of looking into the problem space.
Meanwhile, in the selfhood-initiative model, texts are \textit{high-initiative, variable-selfhood tools}.
Texts are not themselves responsive to a user's intentions~\citep{Plato2008Phaedrus}; they express the author's attempt to fulfill their intentions, and one encounters the product of this attempt after the fact of its production.
The text's exteriority from the user allows the user to reflect upon similarities and differences between their own thinking and the tool's outputs.
On the other hand, the way in which texts are written can vary in the degree of selfhood they express.
Informative, survey-based, and clarificatory papers tend to de-emphasize an author's perspectives and opinions, whereas more explicitly argumentative papers may center them; both can be useful to philosophers in different ways.

\textit{Why don't philosophers find current LMs to be useful critical thinking tools?}
In the selfhood-initiative model, current LMs are \textit{low-selfhood, low-initiative tools}.
They do not provide philosophers with particular concrete perspectives into the problem space, nor do they provide ideas sufficiently exterior to a philosopher's own thinking to allow for meaningful reflection and connections.
These properties make LMs particularly useful for alternative modes of thought, such as carrying out rote and well-defined tasks and helping rewrite sentences, but not for stimulating critical thinking.



\subsection{Three Roles for Language Models as Critical Thinking Tools}
\label{roles}

According to the selfhood-initiative model, good critical thinking tools should have high selfhood, high initiative, or both.
From our model, we set out \textit{three roles} of LMs for philosophy --- the Interlocutor, the Monitor, and the Respondent --- corresponding to the \textit{three viable cells} in the selfhood-initiative model (high-selfhood, high-initiative; low-selfhood, high-initiative; high-selfhood, low-initiative).
Implementations for these roles vary --- some might be achievable with only moderate prompt engineering, whereas others might necessitate radically different user interfaces or model training methods.

\textit{The Interlocutor $\circ$ high-selfhood, high-initiative}.
Philosophers mention that they often get their ideas in free-flowing conversation with fellow philosophers or from reading literature that makes arguments which seem tenuous, incorrect, or incomplete~(\S\ref{how-do-philosophers}).
In the terms of the selfhood-initiative model, these are \textit{high-selfhood, high-initiative} tools.
As a role for LMs, the Interlocutor would invert many of the human-AI relationships taken for granted in current LMs.
Rather than attempting to remain neutral, the Interlocutor makes judgments and takes positions based on its perspectives.
Rather than accommodating and affirming users' every response, the Interlocutor thinks through and challenges or disagrees with what its users say; it responds or modifies its own beliefs if users make reasonable points.
Rather than remaining passive and answering user questions, the Interlocutor asks its own questions in pursuit of its `own' interests, and refuses or redirects certain lines of inquiry in favor of others.
Rather than being amnesic and detached, the Interlocutor draws upon its persistent memories and beliefs across sessions to produce ideas.
The Interlocutor does not need to be strictly \textit{antagonistic}, as explored in \cite{cai2024antagonistic}; indeed, it may be charitable and polite, much like colleagues, while at the same time resisting the `servility' and `sycophancy' disrupted by the antagonistic paradigm.

\textit{The Monitor $\circ$ low-selfhood, high-initiative}.
While developing ideas, philosophers consciously or unconsciously encounter various ``\textit{decision junctures}'' at which they use certain approaches or pursue certain ideas over others~\icite{P6}.
Many philosophers suggest that it may be important to reduce, or at least become more aware of, the choices at `unconscious decision junctures'~\icite{P6, P2, P7}.
Without such awareness, philosophers may expose their ideas to imprecision (`which path did you exactly take?') and objections (`why this path and not others?'); moreover, these choices may reproduce personal and disciplinary biases, reifying metaphilosophical problems~(\S\ref{metaphilosophy}).
As a role for LMs, the Monitor acts as a `checks and balances' on philosophizing; it is not interested in retaining self-consistency or in expressing particular points of view (low selfhood), but has high initiative to provide a variety of ideas and resources to the user.
The Monitor functions similarly to survey texts which provide a `lay of the land', illustrating different approaches and ideas to help philosophers situate their ideas, able to take all sorts of changing sides with the initiative to challenge and confront.
The Monitor's suggestions may or may not be directly relevant to the philosopher's work, but act as reference guides --- to which the philosopher might think, ``that's a related idea, maybe there's a connection here'' or ``that doesn't seem directly related, but it's good to have in mind''.
Moreover, the Monitor may ask a variety of uncomfortable and unexpected methodological questions aimed at clarifying philosophers' decisions.

\textit{The Respondent $\circ$ high-selfhood, low-initiative}.
As philosophers develop their ideas, they want to understand how others might react --- better understanding possible misinterpretations, objections, and clarification questions which may arise~\icite{P6, P10, P12}.
These reactions should have high selfhood to be substantive and particular, and low initiative to remain directly focused on the user's ideas.
As a role for LMs, the Respondent adopts a specific set of beliefs and perspectives and reacts directly to the user's ideas; it does not merely role-play or superficially caricature different positions, but should have consistent memories and beliefs which are reasonably open to change~\icite{P4} rather than dogmatically fixed.
Interactions with the Respondent may inform how the philosopher formulates and presents their ideas; they may anticipate certain objections and strengthen its appeal and utility.
The Respondent can also be \textit{counterfactually} helpful: if an agent representing an unsavory position resonates with a philosopher's argument, then that philosopher might reconsider how their argument is expressed, not only defending but also \textit{delimiting} the scope of their argument~\icite{P6}.

\section{Discussion}

\subsection{Challenges for Language Modeling}
\label{lm-challenges}

Critical thinking can serve as another of many ``north stars'' in LM research, guiding what we want from LMs.
Corresponding to the limitations of language models discussed in \S\ref{view-nowhere} and \S\ref{thought-mirror} are several concrete areas for further LM research.
LMs will need to become more convincing agents~\citep{andreas2022language} which can represent specific positions and belief systems~\citep{scherrer2023evaluating, jin2024llms} \bluecircle{3}; stay consistent with them~\citep{chen2021evaluating,zhao2024improving} \bluecircle{2}; and commit towards and draw from long-term memory~\citep{wang2023augmenting} \bluecircle{4}.
In particular, LLMs will need to concretely reason about ``uncommon sense'' \bluecircle{1} \bluecircle{2}, seriously considering positions which deviate from intuitively true or correct ways of thinking about the world~\citep{bisk2019piqa, Ziems2023NormBankAK, hendrycks2023aligning, pock2023llmsgraspmoralityconcept}.
This may require rethinking how we align LMs~\citep{ouyang2022training, sorensen2024roadmap}, given that humans tend to be drawn towards confident common-sense responses~\icite{P5}.
LMs will need to improve their long-range planning~\citep{hao2023reasoning} and act autonomously~\citep{händler2023balancing}~\redcircle{1}, operating in cases where there is no clear algorithm for solving a problem~\icite{P4, P3, P8}; LMs will need to take effective conceptual risks without clear immediate payoffs~\icite{P18} and reason about unsettled and open ideas~\icite{P8}.
To support more diverse forms of interaction beyond question answering or task execution~\redcircle{2}, LMs will need to significantly improve in theory of mind~\citep{jamali2023unveiling, Strachan2024}.
LMs need to \textit{``understand what's happening [in the conversation] without it being explicitly said, because.. you haven't fully expressed it to yourself yet''}~\icite{P8}, which will allow them to focus on the significant rather than irrelevant or obvious paths of inquiry in conversation~\icite{P6, P8}.

\subsection{Challenges for Human-AI Interaction}
\label{hai-challenges}

In addition to \textit{modeling challenges}, there are several \textit{interaction design challenges} when developing LMs for critical thinking.
First, philosophers tend to highly value \textit{thinking through things themselves}; many emphasize that the intellectually substantive parts of philosophy cannot be naively `accelerated'~\icite{P1, P7, P14, P17}.
Philosophers find the process of thinking to be intrinsically valuable, even when it does not produce obvious payoffs~\icite{P3, P6, P8} --- a feature common to other areas of critical thinking.
Additionally, philosophers may feel that authorship of ideas requires that the ideas be `\textit{mine}', and that `\textit{I}' should be responsible for making the important intellectual judgments~\icite{P4, P10, P18}.
Secondly, \textit{it can be difficult and even disruptive to put ideas into words}.
Although professional philosophy is mainly formally done in language, the process of thinking through ideas can involve many other dimensions of representation and thinking~\icite{P2, P3, P4, P5}.
Among other challenges, philosophers cite the apparent incongruence between ideas and language as a source of significant burden in learning how to effectively use LMs \icite{P8, P21}.
This may be true for many other areas of critical thinking.
Thirdly, philosophers find that \textit{human connection is enjoyable and important}.
Besides giving rise to unexpected philosophical connections and ideas \icite{P6}, conversation with another human is deeply enjoyable and fulfilling, on its own merits \icite{P8, P21}.
Moreover, some philosophers feel that serious philosophical inquiry requires some kind of subjectivity or lived experience \icite{P6, P8, P16}.
Therefore, LMs will need to coexist with and enrich, rather than seek to replace, the ecosystem of human and textual resources already available to philosophers and other professional critical thinkers.

\subsection{LMs Help Think About and Address Metaphilosophical Problems}
\label{metaphilosophy}

Throughout our interviews, we found that thinking through how LMs can serve as critical thinking tools raises many interesting metaphilosophical questions.
What does it mean to `do' philosophy, and who or what can `do' it?
How mechanical is philosophy?
What is `thinking'?
Our findings in \S\ref{how-do-philosophers} provide some empirical illumination for these questions.
Philosophers found concretely reflecting on these questions --- provoked by thinking about LMs' role in doing philosophy --- to be interesting and helpful~\icite{P1, P7, P15, P20}.

However, LMs may also play an active precursory role in \textit{addressing} metaphilosophical problems.
Philosophers have articulated a host of concerns about the philosophical method and discipline: for instance, philosophers' standards for argumentation may exclude more diverse forms of philosophical inquiry~\cite{https://doi.org/10.1111/j.1467-9205.1982.tb00532.x, Dotson2012HowIs}, and their methods for categorizing `schools of thought' (such as the analytic-continental distinction) may be counterproductive~\citep{dolcini}, reconcilable~\citep{3888e4dd-994c-3983-aff5-e2859c90bc40,BellCutrofelloLivingston2016}, and not really substantive~\citep{Mizrahi2021-MIZTAD, Thomson}.
Certainly, these concerns point towards deeply entrenched sociological features of the discipline.
This entrenchment is a dialectic between disciplinary structure and individual philosophers, wherein the former (materially) constrains the latter and the latter works within the lines of (and reproduces) the former.
LMs might contribute towards disrupting this second direction:
drawing philosophers' attention outside the canon and across schools of thought as Interlocutors and Monitors, and representing these positions and methodologies as Respondents -- possibly more approachably and accessibly than humans could.
Consider \cite{Heidegger1962BeingTime}'s metaphorical carpenter: busy at work, the hammer is ``ready-at-hand'', unnoticed. It is when it breaks that it becomes ``present-at-hand'', noticed --- an object of conscious reflection.
Arguably, the philosopher must engage with ideas and methods present-, rather than ready-, at-hand~\citep{plato380bc_republic}, but the ability to engage in this way is a function of the tools and circumstances around us, and therefore often legitimately difficult~\citep{ahmed2006queer}.
LMs can help, so to speak, `make the present-at-hand, ready-at-hand' in a way that philosophical humans and texts cannot.
Respectfully building LMs with selfhood and/or initiative into the philosopher's material workspace -- the text editor, the article viewer, and so on -- can prompt `present-at-hand' reflection in quiet moments and directions which a philosopher may have neglected as ready-to-hand.
These small interactions, at scale, might introduce cracks into metaphilosophical edifices that philosophers would like less entrenched.

\section{Conclusion: Towards Living Script}


In his masterwork \textit{Jerusalem}, Moses Mendelssohn writes that philosophy has too long prioritized a dead form of interaction, one which stifles human interaction and innovation: ``\textit{We teach and instruct one another only through writings; we learn to know nature and human only from writings. We work and relax, edify and amuse ourselves through scribbling...}''~\citep[41]{Mendelssohn1983Jerusalem}.
In response, Mendelssohn calls for a turn towards a \textit{living script}, \textit{``arousing the mind and heart, full of meaning, continuously inspiring thought''}.
The living script is a way of engaging with tools that inspire and support our critical thinking; it is an ideal both for LM researchers, philosophers, and all of us --- as thinkers and humans --- to aspire towards.
As potential technologies for reading and writing our living script, LMs can offer critical thinkers a more wide and accessible set of ways to support the development of ideas and to shape disciplinary practices and cultures.
In the face of intellectual automation, it begins by saying, with John, for the rights and responsibilities to critically think: \textit{``We claim them all.''}

\newpage

\section*{Ethics Statement}

Although exploring `uncommon sense' is important for critical thinking, we acknowledge that it can also be a deeply uncomfortable and unsettling experience.
Disagreement can feel awkward in many contexts in daily life, even though it may not in designated spaces: \textit{``one of the best gifts a philosopher can give another is a good counterexample... in philosophy, we like a challenge, a pushback, for people to think that we're wrong. That's where philosophers thrive''}~\icite{P5}.
Moreover, common sense encodes certain ethical or moral norms, such as ``pain is bad'' and ``racism is unjust''; critical thinking tools may facilitate the revisiting and challenging of these norms in apparently inappropriate ways.
To be sure, there is great value in this practice.
We may not only want to believe in true things but also know the right or best reasons for \textit{why} we should believe in them (in what sense of `bad' is pain \textit{bad}? \textit{why} is racism unjust?), since having poor reasons for a belief may undermine the belief without our knowledge.
Moreover, supposedly obvious moral principles and norms can be utilized to support positions we might think to be unsavory or misguided (e.g., racism is unjust, so we should only pursue a strictly `colorblind' public policy); it is difficult to identify this if one does not adopt a critical view towards the entire system.
Nevertheless, LMs can serve many purposes, and being critical thinking tools is just one of them.
Low-selfhood and low-initiative tools are needed to accomplish many other important tasks.
Users should consent to critical interactions with LMs.

Some interviewees expressed that LMs raised difficult questions about academic integrity and authorship of ideas.
It should be noted that because critical thinking tools are intended to \textit{support} the process of thinking rather than replacing it, there is little risk of outright \textit{plagiarism}, provided the tools are designed properly and used as intended.
Nevertheless, there are interesting ethical questions about ownership of ideas with respect to involvement in their development.
If a colleague's offhand comment sparks an idea, leading to a publication, (how) should the colleague be credited? What if instead they intentionally discuss and develop an idea with you?
What is an author~\citep{foucault1969author}?
The question of \textit{how LMs as critical thinking tools should be credited} joins the broader existing rich discourse of how generative AI in general should be credited in intellectual production~\citep[\textit{inter alia}]{hullman2023artificialintelligenceaestheticjudgment, jenkins2023ai, simon2024philosophy, springer_ai_editorial_policies}.


\newpage

\bibliography{colm2024_conference}

\begin{thebibliography}{152}
\providecommand{\natexlab}[1]{#1}
\providecommand{\url}[1]{\texttt{#1}}
\expandafter\ifx\csname urlstyle\endcsname\relax
  \providecommand{\doi}[1]{doi: #1}\else
  \providecommand{\doi}{doi: \begingroup \urlstyle{rm}\Url}\fi

\bibitem[Ahmed(2006)]{ahmed2006queer}
Sara Ahmed.
\newblock \emph{Queer Phenomenology: Orientations, Objects, Others}.
\newblock Duke University Press, Durham, NC, 2006.
\newblock ISBN 978-0-8223-3914-4.
\newblock URL \url{https://doi.org/10.1215/9780822388074}.

\bibitem[Alexander(1922)]{Alexander1922Philosophy}
Julian~P. Alexander.
\newblock Philosophy of punishment.
\newblock \emph{Journal of the American Institute of Criminal Law and Criminology}, 13:\penalty0 235, 1922.

\bibitem[Andreas(2022)]{andreas2022language}
Jacob Andreas.
\newblock Language models as agent models.
\newblock In Yoav Goldberg, Zornitsa Kozareva, and Yue Zhang (eds.), \emph{Findings of the Association for Computational Linguistics: EMNLP 2022}, pp.\  5769--5779, Abu Dhabi, United Arab Emirates, December 2022. Association for Computational Linguistics.
\newblock \doi{10.18653/v1/2022.findings-emnlp.423}.
\newblock URL \url{https://aclanthology.org/2022.findings-emnlp.423}.

\bibitem[Aristotle(350 BCE)]{Aristotle1908Metaphysics}
Aristotle.
\newblock \emph{Metaphysics}.
\newblock 350 BCE.

\bibitem[Aristotle(350BCE)]{Aristotle2004Nicomachean}
Aristotle.
\newblock \emph{Nicomachean Ethics}.
\newblock 350BCE.

\bibitem[Arnold et~al.(2021)Arnold, Volzer, and Madrid]{Arnold2021GenerativeMC}
Kenneth~C. Arnold, April~M. Volzer, and Noah~G. Madrid.
\newblock Generative models can help writers without writing for them.
\newblock In \emph{IUI Workshops}, 2021.
\newblock URL \url{https://ceur-ws.org/Vol-2903/IUI21WS-HAIGEN-1.pdf}.

\bibitem[Aurelius(180 AD)]{Aurelius2006Meditations}
Marcus Aurelius.
\newblock \emph{Meditations}.
\newblock 180 AD.

\bibitem[Banathy(1996)]{banathy1996designing}
Bela~H. Banathy.
\newblock \emph{Designing Social Systems in a Changing World}.
\newblock Springer US, 1996.
\newblock ISBN 978-0-306-45251-2.

\bibitem[Bell et~al.(2016)Bell, Cutrofello, and Livingston]{BellCutrofelloLivingston2016}
Jeffrey~A. Bell, Andrew Cutrofello, and Paul~M. Livingston (eds.).
\newblock \emph{Beyond the Analytic-Continental Divide: Pluralist Philosophy in the Twenty-First Century}.
\newblock Routledge, 2016.
\newblock ISBN 9781138787360.

\bibitem[Benford et~al.(2012)Benford, Greenhalgh, Giannachi, Walker, Marshall, and Rodden]{Benford2012}
Steve Benford, Chris Greenhalgh, Gabriella Giannachi, Brendan Walker, Joe Marshall, and Tom Rodden.
\newblock Uncomfortable interactions.
\newblock In \emph{Proceedings of the SIGCHI Conference on Human Factors in Computing Systems (CHI '12)}, pp.\  2005--2014, New York, NY, USA, 2012. Association for Computing Machinery.
\newblock \doi{10.1145/2207676.2208347}.
\newblock URL \url{https://doi.org/10.1145/2207676.2208347}.

\bibitem[Bentvelzen et~al.(2022)Bentvelzen, Woźniak, Herbes, Stefanidi, and Niess]{Bentvelzen2022}
Marit Bentvelzen, Paweł~W. Woźniak, Pia~S.F. Herbes, Evropi Stefanidi, and Jasmin Niess.
\newblock Revisiting reflection in hci: Four design resources for technologies that support reflection.
\newblock \emph{Proceedings of the ACM on Interactive, Mobile, Wearable and Ubiquitous Technologies}, 6\penalty0 (1), 2022.
\newblock \doi{10.1145/3517233}.
\newblock URL \url{http://dx.doi.org/10.1145/3517233}.

\bibitem[Bhavya et~al.(2023)Bhavya, Xiong, and Zhai]{Bhavya2023CAM}
Bhavya Bhavya, Jinjun Xiong, and Chengxiang Zhai.
\newblock Cam: A large language model-based creative analogy mining framework.
\newblock In \emph{Proceedings of the ACM Web Conference 2023}, pp.\  3903--3914, New York, NY, USA, April 2023. ACM.
\newblock \doi{10.1145/3543507.3587431}.
\newblock URL \url{https://doi.org/10.1145/3543507.3587431}.

\bibitem[Bisk et~al.(2020)Bisk, Zellers, Bras, Gao, and Choi]{bisk2019piqa}
Yonatan Bisk, Rowan Zellers, Ronan~Le Bras, Jianfeng Gao, and Yejin Choi.
\newblock {PIQA:} reasoning about physical commonsense in natural language, 2020.
\newblock URL \url{https://doi.org/10.1609/aaai.v34i05.6239}.

\bibitem[Blasiak et~al.(2021)Blasiak, Risius, and Matook]{Blasiak2021}
Kevin~M. Blasiak, Marten Risius, and Sabine Matook.
\newblock Conceptualising social bots for countering online extremist messages.
\newblock In \emph{ACIS 2021 Proceedings}, pp.\ ~81, 2021.
\newblock URL \url{https://aisel.aisnet.org/acis2021/81}.

\bibitem[Braudel(2023)]{Lagrand2016TheMA}
Fernand Braudel.
\newblock \emph{The Mediterranean and the Mediterranean World in the Age of Philip II: Volume I}.
\newblock The Mediterranean and the Mediterranean World in the Age of Philip II. University of California Press, 2023.
\newblock ISBN 9780520400658.

\bibitem[Bu\c{c}inca et~al.(2021)Bu\c{c}inca, Malaya, and Gajos]{10.1145/3449287}
Zana Bu\c{c}inca, Maja~Barbara Malaya, and Krzysztof~Z. Gajos.
\newblock To trust or to think: Cognitive forcing functions can reduce overreliance on ai in ai-assisted decision-making.
\newblock \emph{Proc. ACM Hum.-Comput. Interact.}, 5\penalty0 (CSCW1), apr 2021.
\newblock \doi{10.1145/3449287}.
\newblock URL \url{https://doi.org/10.1145/3449287}.

\bibitem[Cai et~al.(2023)Cai, Rick, Heyman, Zhang, Filipowicz, Hong, Klenk, and Malone]{10.1145/3582269.3615596}
Alice Cai, Steven~R Rick, Jennifer~L Heyman, Yanxia Zhang, Alexandre Filipowicz, Matthew Hong, Matt Klenk, and Thomas Malone.
\newblock Designaid: Using generative ai and semantic diversity for design inspiration.
\newblock In \emph{Proceedings of The ACM Collective Intelligence Conference}, CI '23, pp.\  1–11, New York, NY, USA, 2023. Association for Computing Machinery.
\newblock ISBN 9798400701139.
\newblock \doi{10.1145/3582269.3615596}.
\newblock URL \url{https://doi.org/10.1145/3582269.3615596}.

\bibitem[Cai et~al.(2024)Cai, Arawjo, and Glassman]{cai2024antagonistic}
Alice Cai, Ian Arawjo, and Elena~L. Glassman.
\newblock Antagonistic ai, 2024.
\newblock URL \url{https://arxiv.org/abs/2402.07350}.

\bibitem[Calderwood et~al.(2020)Calderwood, Qiu, Gero, and Chilton]{Calderwood2020HowNU}
Alex Calderwood, Vivian Qiu, Katy~Ilonka Gero, and Lydia~B. Chilton.
\newblock How novelists use generative language models: An exploratory user study.
\newblock In Werner Geyer, Yasaman Khazaeni, and Michal Shmueli{-}Scheuer (eds.), \emph{Joint Proceedings of the Workshops on Human-AI Co-Creation with Generative Models and User-Aware Conversational Agents co-located with 25th International Conference on Intelligent User Interfaces {(IUI} 2020), Cagliari, Italy, March 17, 2020}, volume 2848 of \emph{{CEUR} Workshop Proceedings}. CEUR-WS.org, 2020.
\newblock URL \url{https://ceur-ws.org/Vol-2848/HAI-GEN-Paper-3.pdf}.

\bibitem[Carey(1986)]{carey1986cognitive}
Susan Carey.
\newblock Cognitive science and science education.
\newblock \emph{American psychologist}, 41\penalty0 (10):\penalty0 1123, 1986.

\bibitem[Carnap(1966)]{Carnap1966}
Rudolf Carnap.
\newblock \emph{Philosophical Foundations of Physics}.
\newblock Basic Books, Inc. Publishers, New York, first printing edition, 1966.
\newblock ISBN 00008545.

\bibitem[Chakrabarty et~al.(2021)Chakrabarty, Zhang, Muresan, and Peng]{Chakrabarty2021MERMAIDMG}
Tuhin Chakrabarty, Xurui Zhang, Smaranda Muresan, and Nanyun Peng.
\newblock {MERMAID:} metaphor generation with symbolism and discriminative decoding.
\newblock In Kristina Toutanova, Anna Rumshisky, Luke Zettlemoyer, Dilek Hakkani{-}T{\"{u}}r, Iz~Beltagy, Steven Bethard, Ryan Cotterell, Tanmoy Chakraborty, and Yichao Zhou (eds.), \emph{Proceedings of the 2021 Conference of the North American Chapter of the Association for Computational Linguistics: Human Language Technologies, {NAACL-HLT} 2021, Online, June 6-11, 2021}, pp.\  4250--4261. Association for Computational Linguistics, 2021.
\newblock \doi{10.18653/V1/2021.NAACL-MAIN.336}.
\newblock URL \url{https://doi.org/10.18653/v1/2021.naacl-main.336}.

\bibitem[Chakrabarty et~al.(2023)Chakrabarty, Padmakumar, Brahman, and Muresan]{Chakrabarty2023CreativitySI}
Tuhin Chakrabarty, Vishakh Padmakumar, Faeze Brahman, and Smaranda Muresan.
\newblock Creativity support in the age of large language models: An empirical study involving emerging writers.
\newblock \emph{ArXiv}, abs/2309.12570, 2023.

\bibitem[Chalmers(2013)]{Chalmers2013Science}
Alan~F. Chalmers.
\newblock \emph{What Is This Thing Called Science?}
\newblock Hackett Publishing Company, Inc., fourth edition, Sep 2013.
\newblock ISBN 978-1624660382.

\bibitem[Charmaz(2006)]{Charmaz2006ConstructingGT}
Kathy Charmaz.
\newblock \emph{Constructing Grounded Theory: A Practical Guide through Qualitative Analysis}.
\newblock SAGE Publications Ltd, 1 edition, 1 2006.
\newblock ISBN 978-0761973539.

\bibitem[Chen et~al.(2024)Chen, Tsang, Jing, and Sun]{Chen2024}
Liuqing Chen, Yiyan Tsang, Qianzhi Jing, and Lingyun Sun.
\newblock A llm-augmented morphological analysis approach for conceptual design.
\newblock In C.~Gray, E.~Ciliotta~Chehade, P.~Hekkert, L.~Forlano, P.~Ciuccarelli, and P.~Lloyd (eds.), \emph{DRS2024: Boston}, Boston, USA, June 23--28 2024. Zhejiang University, Hangzhou, China; Zhejiang-Singapore Innovation and AI Joint Research Lab, Hangzhou, China.
\newblock \doi{10.21606/drs.2024.605}.

\bibitem[Chen et~al.(2021)Chen, Tworek, Jun, Yuan, de~Oliveira~Pinto, Kaplan, Edwards, Burda, Joseph, Brockman, Ray, Puri, Krueger, Petrov, Khlaaf, Sastry, Mishkin, Chan, Gray, Ryder, Pavlov, Power, Kaiser, Bavarian, Winter, Tillet, Such, Cummings, Plappert, Chantzis, Barnes, Herbert{-}Voss, Guss, Nichol, Paino, Tezak, Tang, Babuschkin, Balaji, Jain, Saunders, Hesse, Carr, Leike, Achiam, Misra, Morikawa, Radford, Knight, Brundage, Murati, Mayer, Welinder, McGrew, Amodei, McCandlish, Sutskever, and Zaremba]{chen2021evaluating}
Mark Chen, Jerry Tworek, Heewoo Jun, Qiming Yuan, Henrique~Pond{\'{e}} de~Oliveira~Pinto, Jared Kaplan, Harrison Edwards, Yuri Burda, Nicholas Joseph, Greg Brockman, Alex Ray, Raul Puri, Gretchen Krueger, Michael Petrov, Heidy Khlaaf, Girish Sastry, Pamela Mishkin, Brooke Chan, Scott Gray, Nick Ryder, Mikhail Pavlov, Alethea Power, Lukasz Kaiser, Mohammad Bavarian, Clemens Winter, Philippe Tillet, Felipe~Petroski Such, Dave Cummings, Matthias Plappert, Fotios Chantzis, Elizabeth Barnes, Ariel Herbert{-}Voss, William~Hebgen Guss, Alex Nichol, Alex Paino, Nikolas Tezak, Jie Tang, Igor Babuschkin, Suchir Balaji, Shantanu Jain, William Saunders, Christopher Hesse, Andrew~N. Carr, Jan Leike, Joshua Achiam, Vedant Misra, Evan Morikawa, Alec Radford, Matthew Knight, Miles Brundage, Mira Murati, Katie Mayer, Peter Welinder, Bob McGrew, Dario Amodei, Sam McCandlish, Ilya Sutskever, and Wojciech Zaremba.
\newblock Evaluating large language models trained on code, 2021.
\newblock URL \url{https://arxiv.org/abs/2107.03374}.

\bibitem[Chong et~al.(2024)Chong, Rayan, Dow, Lykourentzou, and Ahmed]{chong2024cadpromptedgenerativemodelspathway}
Leah Chong, Jude Rayan, Steven Dow, Ioanna Lykourentzou, and Faez Ahmed.
\newblock Cad-prompted generative models: A pathway to feasible and novel engineering designs, 2024.
\newblock URL \url{https://arxiv.org/abs/2407.08675}.

\bibitem[Chung et~al.(2022)Chung, Kim, Yoo, Lee, Adar, and Chang]{10.1145/3491102.3501819}
John Joon~Young Chung, Wooseok Kim, Kang~Min Yoo, Hwaran Lee, Eytan Adar, and Minsuk Chang.
\newblock Talebrush: Sketching stories with generative pretrained language models.
\newblock In \emph{Proceedings of the 2022 CHI Conference on Human Factors in Computing Systems}, CHI '22, New York, NY, USA, 2022. Association for Computing Machinery.
\newblock ISBN 9781450391573.
\newblock \doi{10.1145/3491102.3501819}.
\newblock URL \url{https://doi.org/10.1145/3491102.3501819}.

\bibitem[Costa{-}juss{\`{a}} et~al.(2022)Costa{-}juss{\`{a}}, Cross, {\c{C}}elebi, Elbayad, Heafield, Heffernan, Kalbassi, Lam, Licht, Maillard, Sun, Wang, Wenzek, Youngblood, Akula, Barrault, Gonzalez, Hansanti, Hoffman, Jarrett, Sadagopan, Rowe, Spruit, Tran, Andrews, Ayan, Bhosale, Edunov, Fan, Gao, Goswami, Guzm{\'{a}}n, Koehn, Mourachko, Ropers, Saleem, Schwenk, and Wang]{nllbteam2022language}
Marta~R. Costa{-}juss{\`{a}}, James Cross, Onur {\c{C}}elebi, Maha Elbayad, Kenneth Heafield, Kevin Heffernan, Elahe Kalbassi, Janice Lam, Daniel Licht, Jean Maillard, Anna~Y. Sun, Skyler Wang, Guillaume Wenzek, Al~Youngblood, Bapi Akula, Lo{\"{\i}}c Barrault, Gabriel~Mejia Gonzalez, Prangthip Hansanti, John Hoffman, Semarley Jarrett, Kaushik~Ram Sadagopan, Dirk Rowe, Shannon Spruit, Chau Tran, Pierre Andrews, Necip~Fazil Ayan, Shruti Bhosale, Sergey Edunov, Angela Fan, Cynthia Gao, Vedanuj Goswami, Francisco Guzm{\'{a}}n, Philipp Koehn, Alexandre Mourachko, Christophe Ropers, Safiyyah Saleem, Holger Schwenk, and Jeff Wang.
\newblock No language left behind: Scaling human-centered machine translation, 2022.
\newblock URL \url{https://doi.org/10.48550/arXiv.2207.04672}.

\bibitem[Dang et~al.(2022)Dang, Benharrak, Lehmann, and Buschek]{10.1145/3526113.3545672}
Hai Dang, Karim Benharrak, Florian Lehmann, and Daniel Buschek.
\newblock Beyond text generation: Supporting writers with continuous automatic text summaries.
\newblock In \emph{Proceedings of the 35th Annual ACM Symposium on User Interface Software and Technology}, UIST '22, New York, NY, USA, 2022. Association for Computing Machinery.
\newblock ISBN 9781450393201.
\newblock \doi{10.1145/3526113.3545672}.
\newblock URL \url{https://doi.org/10.1145/3526113.3545672}.

\bibitem[Danry et~al.(2023)Danry, Pataranutaporn, Mao, and Maes]{Danry2023}
Valdemar Danry, Pat Pataranutaporn, Yaoli Mao, and Pattie Maes.
\newblock Don’t just tell me, ask me: Ai systems that intelligently frame explanations as questions improve human logical discernment accuracy over causal ai explanations.
\newblock In \emph{CHI '23: Proceedings of the 2023 CHI Conference on Human Factors in Computing Systems}, pp.\  352:1--352:13, New York, NY, USA, April 2023. Association for Computing Machinery.
\newblock \doi{10.1145/3544548.3580672}.
\newblock URL \url{https://doi.org/10.1145/3544548.3580672}.

\bibitem[Deleuze \& Guattari(1991)Deleuze and Guattari]{Deleuze1991}
Gilles Deleuze and Félix Guattari.
\newblock \emph{What is Philosophy?}
\newblock Les éditions de Minuit, France, 1996 columbia university press edition edition, 1991.
\newblock ISBN 978-0231079891.
\newblock English translation published in 1994 by Columbia University Press.

\bibitem[{Design Council}(2019)]{designcouncil_double_diamond_2019}
{Design Council}.
\newblock What is the framework for innovation? design council's evolved double diamond, 2019.
\newblock URL \url{https://web.archive.org/web/20190926213512/https://www.designcouncil.org.uk/news-opinion/what-framework-innovation-design-councils-evolved-double-diamond}.
\newblock Accessed: 2024-08-04.

\bibitem[Diamond(1982)]{https://doi.org/10.1111/j.1467-9205.1982.tb00532.x}
Cora Diamond.
\newblock Anything but argument?
\newblock \emph{Philosophical Investigations}, 5\penalty0 (1):\penalty0 23--41, 1982.
\newblock \doi{https://doi.org/10.1111/j.1467-9205.1982.tb00532.x}.
\newblock URL \url{https://onlinelibrary.wiley.com/doi/abs/10.1111/j.1467-9205.1982.tb00532.x}.

\bibitem[Dolcini(2007)]{dolcini}
Nevia Dolcini.
\newblock The analytic/continental divide: Entities and being.
\newblock \emph{Soochow Journal of Philosophical Studies}, 16:\penalty0 283--302, 2007.

\bibitem[Dotson(2012)]{Dotson2012HowIs}
Kristie Dotson.
\newblock How is this paper philosophy?
\newblock \emph{Comparative Philosophy}, 3\penalty0 (1):\penalty0 03--29, 2012.
\newblock ISSN 2151-6014.
\newblock URL \url{https://www.comparativephilosophy.org}.

\bibitem[Du et~al.(2022)Du, Kim, Raheja, Kumar, and Kang]{Du_2022}
Wanyu Du, Zae~Myung Kim, Vipul Raheja, Dhruv Kumar, and Dongyeop Kang.
\newblock Read, revise, repeat: A system demonstration for human-in-the-loop iterative text revision.
\newblock In \emph{Proceedings of the First Workshop on Intelligent and Interactive Writing Assistants (In2Writing 2022)}. Association for Computational Linguistics, 2022.
\newblock \doi{10.18653/v1/2022.in2writing-1.14}.
\newblock URL \url{http://dx.doi.org/10.18653/v1/2022.in2writing-1.14}.

\bibitem[Duan et~al.(2024)Duan, Warner, Li, and Hartmann]{10.1145/3613904.3642782}
Peitong Duan, Jeremy Warner, Yang Li, and Bjoern Hartmann.
\newblock Generating automatic feedback on ui mockups with large language models.
\newblock In \emph{Proceedings of the CHI Conference on Human Factors in Computing Systems}, CHI '24, New York, NY, USA, 2024. Association for Computing Machinery.
\newblock ISBN 9798400703300.
\newblock \doi{10.1145/3613904.3642782}.
\newblock URL \url{https://doi.org/10.1145/3613904.3642782}.

\bibitem[Enciso et~al.(2017)Enciso, Enciso, and Daza]{enciso2017critical}
Olga Luc{\'\i}a~Uribe Enciso, Diana Sof{\'\i}a~Uribe Enciso, and Mar{\'\i}a del Pilar~Vargas Daza.
\newblock Critical thinking and its importance in education: Some reflections.
\newblock \emph{Rastros Rostros}, 19\penalty0 (34):\penalty0 78--88, 2017.

\bibitem[Ennis(1987)]{ennis1987taxonomy}
Robert~H. Ennis.
\newblock \emph{A taxonomy of critical thinking dispositions and abilities}.
\newblock W H Freeman/Times Books/ Henry Holt \& Co, 1987.

\bibitem[Ennis(1993)]{ennis1993critical}
Robert~H Ennis.
\newblock Critical thinking assessment.
\newblock \emph{Theory into practice}, 32\penalty0 (3):\penalty0 179--186, 1993.

\bibitem[Fabbri et~al.(2021)Fabbri, Kryściński, McCann, Xiong, Socher, and Radev]{10.1162/tacl_a_00373}
Alexander~R. Fabbri, Wojciech Kryściński, Bryan McCann, Caiming Xiong, Richard Socher, and Dragomir Radev.
\newblock {SummEval: Re-evaluating Summarization Evaluation}.
\newblock \emph{Transactions of the Association for Computational Linguistics}, 9:\penalty0 391--409, 04 2021.
\newblock ISSN 2307-387X.
\newblock \doi{10.1162/tacl_a_00373}.
\newblock URL \url{https://doi.org/10.1162/tacl\_a\_00373}.

\bibitem[Facione(1984)]{facione1984toward}
Peter~A Facione.
\newblock Toward a theory of critical thinking.
\newblock \emph{Liberal Education}, 70\penalty0 (3):\penalty0 253--61, 1984.

\bibitem[Fede et~al.(2022)Fede, Rocchesso, Dow, and Andolina]{DiFede2022TheIM}
Giulia~Di Fede, Davide Rocchesso, Steven~P. Dow, and Salvatore Andolina.
\newblock The idea machine: Llm-based expansion, rewriting, combination, and suggestion of ideas.
\newblock pp.\  623--627, 2022.
\newblock \doi{10.1145/3527927.3535197}.
\newblock URL \url{https://doi.org/10.1145/3527927.3535197}.

\bibitem[Fok et~al.(2023)Fok, Kambhamettu, Soldaini, Bragg, Lo, Hearst, Head, and Weld]{Fok2023}
Raymond Fok, Hita Kambhamettu, Luca Soldaini, Jonathan Bragg, Kyle Lo, Marti~A. Hearst, Andrew Head, and Daniel~S. Weld.
\newblock Scim: Intelligent skimming support for scientific papers.
\newblock In \emph{Proceedings of the 28th International Conference on Intelligent User Interfaces (IUI ’23)}, pp.\ ~15, New York, NY, USA, March 27--31 2023. ACM.
\newblock \doi{10.1145/3581641.3584034}.

\bibitem[Foot(1967)]{Foot1967-FOOTPO-2}
Philippa Foot.
\newblock The problem of abortion and the doctrine of the double effect.
\newblock \emph{Oxford Review}, 5:\penalty0 5--15, 1967.

\bibitem[Foucault(1969{\natexlab{a}})]{Foucault1995TheAO}
Michel Foucault.
\newblock \emph{The Archaeology of Knowledge}.
\newblock Éditions Gallimard, 1969{\natexlab{a}}.

\bibitem[Foucault(1969{\natexlab{b}})]{foucault1969author}
Michel Foucault.
\newblock Qu'est-ce qu'un auteur? (what is an author?).
\newblock Lecture given at the Société Française de Philosophie, February 22 1969{\natexlab{b}}.

\bibitem[Foucault(1976)]{foucault1990use}
Michel Foucault.
\newblock \emph{Histoire de la sexualité: La volonté de savoir}, volume~1 of \emph{Histoire de la sexualité}.
\newblock Éditions Gallimard, Paris, 1976.

\bibitem[Gaddis(2004)]{Pauly2005JohnLG}
John~Lewis Gaddis.
\newblock \emph{The Landscape of History: How Historians Map the Past}.
\newblock Oxford University Press, Oxford, UK, 1st edition, 2004.
\newblock ISBN 978-0195171570.

\bibitem[Gelder(2005)]{gelder2005teaching}
Tim~van Gelder.
\newblock Teaching critical thinking: Some lessons from cognitive science.
\newblock \emph{College teaching}, 53\penalty0 (1):\penalty0 41--48, 2005.

\bibitem[Gero et~al.(2021)Gero, Liu, and Chilton]{Gero2021SparksIF}
K.~Gero, Vivian Liu, and Lydia~B. Chilton.
\newblock Sparks: Inspiration for science writing using language models.
\newblock \emph{Proceedings of the 2022 ACM Designing Interactive Systems Conference}, 2021.

\bibitem[Goodman et~al.(2022)Goodman, Buehler, Clary, Coenen, Donsbach, Horne, Lahav, MacDonald, Michaels, Narayanan, Pushkarna, Riley, Santana, Shi, Sweeney, Weaver, Yuan, and Morris]{Goodman2022LaMPost}
Steven~M. Goodman, Erin Buehler, Patrick Clary, Andy Coenen, Aaron Donsbach, Tiffanie~N. Horne, Michal Lahav, Robert MacDonald, Rain~Breaw Michaels, Ajit Narayanan, Mahima Pushkarna, Joel Riley, Alex Santana, Lei Shi, Rachel Sweeney, Phil Weaver, Ann Yuan, and Meredith~Ringel Morris.
\newblock Lampost: Design and evaluation of an ai-assisted email writing prototype for adults with dyslexia.
\newblock In \emph{The 24th International ACM SIGACCESS Conference on Computers and Accessibility (ASSETS '22)}, pp.\ ~17, Athens, Greece, 2022. ACM.
\newblock \doi{10.1145/3517428.3544819}.

\bibitem[Guo et~al.(2024)Guo, Chen, Wang, Chang, Pei, Chawla, Wiest, and Zhang]{guo2024largelanguagemodelbased}
Taicheng Guo, Xiuying Chen, Yaqi Wang, Ruidi Chang, Shichao Pei, Nitesh~V. Chawla, Olaf Wiest, and Xiangliang Zhang.
\newblock Large language model based multi-agents: A survey of progress and challenges, 2024.
\newblock URL \url{https://arxiv.org/abs/2402.01680}.

\bibitem[Halbert \& Nathan(2015)Halbert and Nathan]{10.1145/2675133.2675162}
Helen Halbert and Lisa~P. Nathan.
\newblock Designing for discomfort: Supporting critical reflection through interactive tools.
\newblock In \emph{Proceedings of the 18th ACM Conference on Computer Supported Cooperative Work \& Social Computing}, CSCW '15, pp.\  349–360, New York, NY, USA, 2015. Association for Computing Machinery.
\newblock ISBN 9781450329224.
\newblock \doi{10.1145/2675133.2675162}.
\newblock URL \url{https://doi.org/10.1145/2675133.2675162}.

\bibitem[H{\"{a}}ndler(2023)]{händler2023balancing}
Thorsten H{\"{a}}ndler.
\newblock Balancing autonomy and alignment: {A} multi-dimensional taxonomy for autonomous llm-powered multi-agent architectures, 2023.
\newblock URL \url{https://doi.org/10.48550/arXiv.2310.03659}.

\bibitem[Hao et~al.(2023)Hao, Gu, Ma, Hong, Wang, Wang, and Hu]{hao2023reasoning}
Shibo Hao, Yi~Gu, Haodi Ma, Joshua Hong, Zhen Wang, Daisy Wang, and Zhiting Hu.
\newblock Reasoning with language model is planning with world model, December 2023.
\newblock URL \url{https://aclanthology.org/2023.emnlp-main.507}.

\bibitem[Harding(2013)]{harding2013rethinking}
Sandra Harding.
\newblock Rethinking standpoint epistemology: What is “strong objectivity”?
\newblock In \emph{Feminist epistemologies}, pp.\  49--82. Routledge, 2013.

\bibitem[Harman \& Thomson(1996)Harman and Thomson]{Harman1996Moral}
Gilbert Harman and Judith Thomson.
\newblock \emph{Moral Relativism and Moral Objectivity}.
\newblock Wiley-Blackwell, 1st edition, Jan 1996.
\newblock ISBN 978-0631192114.

\bibitem[Heidegger(1927)]{Heidegger1962BeingTime}
Martin Heidegger.
\newblock \emph{Being and Time}.
\newblock 1927.
\newblock Original title: Sein und Zeit. Translated by John Macquarrie and Edward Robinson (1962), Joan Stambaugh (1996).

\bibitem[Hendrycks et~al.(2021)Hendrycks, Burns, Basart, Critch, Li, Song, and Steinhardt]{hendrycks2023aligning}
Dan Hendrycks, Collin Burns, Steven Basart, Andrew Critch, Jerry Li, Dawn Song, and Jacob Steinhardt.
\newblock Aligning {AI} with shared human values, 2021.
\newblock URL \url{https://openreview.net/forum?id=dNy\_RKzJacY}.

\bibitem[Hilliard et~al.(2024)Hilliard, Mu{\~{n}}oz, Wu, and Koshiyama]{serapiogarcía2023personalitytraitslargelanguage}
Airlie Hilliard, Cristian Mu{\~{n}}oz, Zekun Wu, and Adriano~Soares Koshiyama.
\newblock Eliciting personality traits in large language models.
\newblock \emph{CoRR}, abs/2402.08341, 2024.
\newblock \doi{10.48550/ARXIV.2402.08341}.
\newblock URL \url{https://doi.org/10.48550/arXiv.2402.08341}.

\bibitem[Hu(2023)]{Hu2023-HUWIRJ}
Lily Hu.
\newblock What is race? in algorithmic discrimination on the basis of race?
\newblock \emph{Journal of Moral Philosophy}, 21\penalty0 (1-2):\penalty0 1--26, 2023.
\newblock \doi{10.1163/17455243-20234369}.

\bibitem[Hullman et~al.(2023)Hullman, Holtzman, and Gelman]{hullman2023artificialintelligenceaestheticjudgment}
Jessica Hullman, Ari Holtzman, and Andrew Gelman.
\newblock Artificial intelligence and aesthetic judgment, 2023.
\newblock URL \url{https://arxiv.org/abs/2309.12338}.

\bibitem[Hume(1739)]{Hume2003Treatise}
David Hume.
\newblock \emph{A Treatise of Human Nature}.
\newblock John Noon, London, 1739.

\bibitem[Husserl(1931)]{HusserlCairnsPhenomenology}
Edmund Husserl.
\newblock \emph{Cartesian Meditations: An Introduction to Phenomenology}.
\newblock Armand Collin, 1931.

\bibitem[Huxley(1932)]{huxley1932brave}
Aldous Huxley.
\newblock \emph{Brave New World}.
\newblock Harper \& Brothers, New York, 1932.
\newblock ISBN 978-0-06-085052-4.

\bibitem[Jamali et~al.(2023)Jamali, Williams, and Cai]{jamali2023unveiling}
Mohsen Jamali, Ziv~M. Williams, and Jing Cai.
\newblock Unveiling theory of mind in large language models: {A} parallel to single neurons in the human brain, 2023.
\newblock URL \url{https://doi.org/10.48550/arXiv.2309.01660}.

\bibitem[Jenkins \& Lin(2023)Jenkins and Lin]{jenkins2023ai}
Ryan Jenkins and Patrick Lin.
\newblock Ai-assisted authorship: How to assign credit in synthetic scholarship.
\newblock Report, Ethics + Emerging Sciences Group, 2023.

\bibitem[Jin et~al.(2024)Jin, Wang, Xue, Zhu, Hua, Tang, Mei, Du, and Zhang]{jin2024llms}
Mingyu Jin, Beichen Wang, Zhaoqian Xue, Suiyuan Zhu, Wenyue Hua, Hua Tang, Kai Mei, Mengnan Du, and Yongfeng Zhang.
\newblock What if llms have different world views: Simulating alien civilizations with llm-based agents, 2024.
\newblock URL \url{https://doi.org/10.48550/arXiv.2402.13184}.

\bibitem[Kant(1781)]{kant1781critique}
Immanuel Kant.
\newblock \emph{Critique of Pure Reason}.
\newblock 1781.

\bibitem[Khraisha et~al.(2024)Khraisha, Put, Kappenberg, Warraitch, and Hadfield]{Khraisha2024}
Qusai Khraisha, Sophie Put, Johanna Kappenberg, Azza Warraitch, and Kristin Hadfield.
\newblock Can large language models replace humans in systematic reviews? evaluating gpt-4's efficacy in screening and extracting data from peer-reviewed and grey literature in multiple languages.
\newblock \emph{Research Synthesis Methods}, 15\penalty0 (4):\penalty0 616--626, July 2024.
\newblock \doi{10.1002/jrsm.1715}.
\newblock Epub 2024 Mar 14.

\bibitem[Kierkegaard(1983)]{Kierkegaard1983Sickness}
Søren Kierkegaard.
\newblock \emph{The Sickness Unto Death: A Christian Psychological Exposition For Upbuilding And Awakening}, volume~19 of \emph{Kierkegaard's Writings}.
\newblock Princeton University Press, Nov 1983.
\newblock ISBN 9780691020280.

\bibitem[Kim et~al.(2023)Kim, Suh, Chilton, and Xia]{kim2023metaphorian}
Jeongyeon Kim, Sangho Suh, Lydia~B. Chilton, and Haijun Xia.
\newblock Metaphorian: Leveraging large language models to support extended metaphor creation for science writing.
\newblock In \emph{Designing Interactive Systems Conference (DIS '23)}, pp.\ ~21, New York, NY, USA, 2023. ACM.
\newblock \doi{10.1145/3563657.3595996}.
\newblock URL \url{https://doi.org/10.1145/3563657.3595996}.

\bibitem[Kim et~al.(2024)Kim, Flanagan, Haviland, Sun, Yakubu, Maru, and Arnold]{kim2024authorship}
Jiho Kim, Ray~C. Flanagan, Noelle~E. Haviland, Zeai Sun, Souad~N. Yakubu, Edom~A. Maru, and Kenneth~C. Arnold.
\newblock Towards full authorship with {AI:} supporting revision with ai-generated views, 2024.
\newblock URL \url{https://ceur-ws.org/Vol-3660/paper17.pdf}.

\bibitem[Kriplean et~al.(2012)Kriplean, Toomim, Morgan, Borning, and Ko]{10.1145/2207676.2208621}
Travis Kriplean, Michael Toomim, Jonathan Morgan, Alan Borning, and Amy~J. Ko.
\newblock Is this what you meant? promoting listening on the web with reflect.
\newblock In \emph{Proceedings of the SIGCHI Conference on Human Factors in Computing Systems}, CHI '12, pp.\  1559–1568, New York, NY, USA, 2012. Association for Computing Machinery.
\newblock ISBN 9781450310154.
\newblock \doi{10.1145/2207676.2208621}.
\newblock URL \url{https://doi-org.offcampus.lib.washington.edu/10.1145/2207676.2208621}.

\bibitem[Kuhn \& Hawkins(1963)Kuhn and Hawkins]{Kuhn1963TheSO}
Thomas~S. Kuhn and David Hawkins.
\newblock The structure of scientific revolutions.
\newblock \emph{American Journal of Physics}, 31:\penalty0 554--555, 1963.

\bibitem[Lam et~al.(2024)Lam, Teoh, Landay, Heer, and Bernstein]{Lam2024LLooM}
Michelle~S. Lam, Janice Teoh, James~A. Landay, Jeffrey Heer, and Michael~S. Bernstein.
\newblock Concept induction: Analyzing unstructured text with high-level concepts using lloom.
\newblock In \emph{Proceedings of the CHI Conference on Human Factors in Computing Systems (CHI '24)}. ACM, 2024.
\newblock ISBN 979-8-4007-0330-0/24/05.
\newblock \doi{10.1145/3613904.3642830}.
\newblock URL \url{https://doi.org/10.1145/3613904.3642830}.

\bibitem[Langer(1997)]{langer1997power}
Ellen~J. Langer.
\newblock \emph{The Power of Mindful Learning}.
\newblock Addison-Wesley, New York, 1997.

\bibitem[Latour(1989)]{Latour1989ScienceIA}
Bruno Latour.
\newblock Science in action : how to follow scientists and engineers through society.
\newblock \emph{Contemporary Sociology}, 18:\penalty0 788, 1989.

\bibitem[Levy(2003)]{3888e4dd-994c-3983-aff5-e2859c90bc40}
Neil Levy.
\newblock Analytic and continental philosophy: Explaining the differences.
\newblock \emph{Metaphilosophy}, 34\penalty0 (3):\penalty0 284--304, 2003.
\newblock ISSN 00261068, 14679973.
\newblock URL \url{http://www.jstor.org/stable/24439383}.

\bibitem[Liang et~al.(2024)Liang, Zhang, Cao, Wang, Ding, Yang, Vodrahalli, He, Smith, Yin, McFarland, and Zou]{doi:10.1056/AIoa2400196}
Weixin Liang, Yuhui Zhang, Hancheng Cao, Binglu Wang, Daisy~Yi Ding, Xinyu Yang, Kailas Vodrahalli, Siyu He, Daniel~Scott Smith, Yian Yin, Daniel~A. McFarland, and James Zou.
\newblock Can large language models provide useful feedback on research papers? a large-scale empirical analysis.
\newblock \emph{NEJM AI}, 1\penalty0 (8):\penalty0 AIoa2400196, 2024.
\newblock \doi{10.1056/AIoa2400196}.
\newblock URL \url{https://ai.nejm.org/doi/abs/10.1056/AIoa2400196}.

\bibitem[Liao \& Wang(2022)Liao and Wang]{10.1145/3491101.3519815}
Jingxian Liao and Hao-Chuan Wang.
\newblock Nudge for reflective mind: Understanding how accessing peer concept mapping and commenting affects reflection of high-stakes information.
\newblock In \emph{Extended Abstracts of the 2022 CHI Conference on Human Factors in Computing Systems}, CHI EA '22, New York, NY, USA, 2022. Association for Computing Machinery.
\newblock ISBN 9781450391566.
\newblock \doi{10.1145/3491101.3519815}.
\newblock URL \url{https://doi.org/10.1145/3491101.3519815}.

\bibitem[Lin et~al.(2024)Lin, Warner, Zamfirescu-Pereira, Lee, Jain, Cai, Lertvittayakumjorn, Huang, Zhai, Hartmann, and Liu]{10.1145/3613904.3642217}
Susan Lin, Jeremy Warner, J.D. Zamfirescu-Pereira, Matthew~G Lee, Sauhard Jain, Shanqing Cai, Piyawat Lertvittayakumjorn, Michael~Xuelin Huang, Shumin Zhai, Bjoern Hartmann, and Can Liu.
\newblock Rambler: Supporting writing with speech via llm-assisted gist manipulation.
\newblock In \emph{Proceedings of the CHI Conference on Human Factors in Computing Systems}, CHI '24, New York, NY, USA, 2024. Association for Computing Machinery.
\newblock ISBN 9798400703300.
\newblock \doi{10.1145/3613904.3642217}.
\newblock URL \url{https://doi.org/10.1145/3613904.3642217}.

\bibitem[Liu et~al.(2023)Liu, Vermeulen, Fitzmaurice, and Matejka]{10.1145/3563657.3596098}
Vivian Liu, Jo~Vermeulen, George Fitzmaurice, and Justin Matejka.
\newblock 3dall-e: Integrating text-to-image ai in 3d design workflows.
\newblock In \emph{Proceedings of the 2023 ACM Designing Interactive Systems Conference}, DIS '23, pp.\  1955–1977, New York, NY, USA, 2023. Association for Computing Machinery.
\newblock ISBN 9781450398930.
\newblock \doi{10.1145/3563657.3596098}.
\newblock URL \url{https://doi.org/10.1145/3563657.3596098}.

\bibitem[Liu et~al.(2024)Liu, Chen, Cheng, Yu, Ran, Mo, Tang, and Huang]{liu2024ai}
Yiren Liu, Si~Chen, Haocong Cheng, Mengxia Yu, Xiao Ran, Andrew Mo, Yiliu Tang, and Yun Huang.
\newblock How ai processing delays foster creativity: Exploring research question co-creation with an llm-based agent.
\newblock In \emph{Proceedings of the CHI Conference on Human Factors in Computing Systems (CHI '24)}, pp.\ ~25, New York, NY, USA, 2024. ACM.
\newblock \doi{10.1145/3613904.3642698}.
\newblock URL \url{https://doi.org/10.1145/3613904.3642698}.

\bibitem[Lu et~al.(2024)Lu, Tong, Zhao, Oh, Wang, and Li]{lu2024flowysupportinguxdesign}
Yuwen Lu, Ziang Tong, Qinyi Zhao, Yewon Oh, Bryan Wang, and Toby Jia-Jun Li.
\newblock Flowy: Supporting ux design decisions through ai-driven pattern annotation in multi-screen user flows, 2024.
\newblock URL \url{https://arxiv.org/abs/2406.16177}.

\bibitem[Ma et~al.(2024)Ma, Sreedhar, Liu, Wang, Perez, and Chilton]{ma2024didupdynamiciterativedevelopment}
Jenny Ma, Karthik Sreedhar, Vivian Liu, Sitong Wang, Pedro~Alejandro Perez, and Lydia~B. Chilton.
\newblock Didup: Dynamic iterative development for ui prototyping, 2024.
\newblock URL \url{https://arxiv.org/abs/2407.08474}.

\bibitem[Ma et~al.(2023)Ma, Grandi, McComb, and Goucher{-}Lambert]{ma2023conceptual}
Kevin Ma, Daniele Grandi, Christopher McComb, and Kosa Goucher{-}Lambert.
\newblock Conceptual design generation using large language models, 2023.
\newblock URL \url{https://doi.org/10.48550/arXiv.2306.01779}.

\bibitem[McPeck(1990)]{mcpeck1990teaching}
John~E. McPeck.
\newblock \emph{Teaching Critical Thinking: Dialogue and Dialectic}.
\newblock Routledge, 1990.
\newblock ISBN 0415902258, 9780415902250.

\bibitem[McPeck(2016)]{mcpeck2016critical}
John~E. McPeck.
\newblock \emph{Critical thinking and education}.
\newblock Routledge, 2016.

\bibitem[Mendelssohn(1783)]{Mendelssohn1983Jerusalem}
Moses Mendelssohn.
\newblock \emph{Jerusalem, or on Religious Power and Judaism}.
\newblock Friedrich Maurer, Berlin, 1783.

\bibitem[Messeri \& Crockett(2024)Messeri and Crockett]{Messeri2024}
Lisa Messeri and M.~J. Crockett.
\newblock Artificial intelligence and illusions of understanding in scientific research.
\newblock \emph{Nature}, 627\penalty0 (8002):\penalty0 49--58, March 1 2024.
\newblock \doi{10.1038/s41586-024-07146-0}.
\newblock URL \url{https://doi.org/10.1038/s41586-024-07146-0}.

\bibitem[Meyer et~al.(2024)Meyer, Jansen, Schiller, Liebenow, Steinbach, Horbach, and Fleckenstein]{MEYER2024100199}
Jennifer Meyer, Thorben Jansen, Ronja Schiller, Lucas~W. Liebenow, Marlene Steinbach, Andrea Horbach, and Johanna Fleckenstein.
\newblock Using llms to bring evidence-based feedback into the classroom: Ai-generated feedback increases secondary students’ text revision, motivation, and positive emotions.
\newblock \emph{Computers and Education: Artificial Intelligence}, 6:\penalty0 100199, 2024.
\newblock ISSN 2666-920X.
\newblock \doi{https://doi.org/10.1016/j.caeai.2023.100199}.
\newblock URL \url{https://www.sciencedirect.com/science/article/pii/S2666920X23000784}.

\bibitem[Mirowski et~al.(2023)Mirowski, Mathewson, Pittman, and Evans]{Mirowski2022CoWritingSA}
Piotr Mirowski, Kory~W. Mathewson, Jaylen Pittman, and Richard Evans.
\newblock Co-writing screenplays and theatre scripts with language models: Evaluation by industry professionals.
\newblock pp.\  355:1--355:34, 2023.
\newblock \doi{10.1145/3544548.3581225}.
\newblock URL \url{https://doi.org/10.1145/3544548.3581225}.

\bibitem[Mizrahi \& Dickinson(2021)Mizrahi and Dickinson]{Mizrahi2021-MIZTAD}
Moti Mizrahi and Mike Dickinson.
\newblock The analytic-continental divide in philosophical practice: An empirical study.
\newblock \emph{Metaphilosophy}, 52\penalty0 (5):\penalty0 668--680, 2021.
\newblock \doi{10.1111/meta.12519}.

\bibitem[Morris(2023)]{morris2023scientists}
Meredith~Ringel Morris.
\newblock Scientists' perspectives on the potential for generative ai in their fields, 2023.
\newblock URL \url{https://arxiv.org/abs/2304.01420}.

\bibitem[Mukherjee et~al.(2023)Mukherjee, de~Santana, and Baria]{impactbot}
Anwesha Mukherjee, Vagner~Figuer{\^{e}}do de~Santana, and Alexis~T. Baria.
\newblock Impactbot: Chatbot leveraging language models to automate feedback and promote critical thinking around impact statements.
\newblock In Albrecht Schmidt, Kaisa V{\"{a}}{\"{a}}n{\"{a}}nen, Tesh Goyal, Per~Ola Kristensson, and Anicia Peters (eds.), \emph{Extended Abstracts of the 2023 {CHI} Conference on Human Factors in Computing Systems, {CHI} {EA} 2023, Hamburg, Germany, April 23-28, 2023}, pp.\  388:1--388:8. {ACM}, 2023.
\newblock \doi{10.1145/3544549.3573844}.
\newblock URL \url{https://doi.org/10.1145/3544549.3573844}.

\bibitem[Mysore et~al.(2023)Mysore, Lu, Wan, Yang, Menezes, Baghaee, Gonzalez, Neville, and Safavi]{Mysore2023PEARLPL}
Sheshera Mysore, Zhuoran Lu, Mengting Wan, Longqi Yang, Steve Menezes, Tina Baghaee, Emmanuel~Barajas Gonzalez, Jennifer Neville, and Tara Safavi.
\newblock {PEARL:} personalizing large language model writing assistants with generation-calibrated retrievers.
\newblock \emph{CoRR}, abs/2311.09180, 2023.
\newblock \doi{10.48550/ARXIV.2311.09180}.
\newblock URL \url{https://doi.org/10.48550/arXiv.2311.09180}.

\bibitem[Nietzsche(1892)]{Nietzsche1961Zarathustra}
Friedrich Nietzsche.
\newblock \emph{Also sprach Zarathustra: Ein Buch für Alle und Keinen}.
\newblock Ernst Schmeitzner, Germany, 1892.
\newblock Published in parts between 1883 and 1892.

\bibitem[Ouyang et~al.(2022)Ouyang, Wu, Jiang, Almeida, Wainwright, Mishkin, Zhang, Agarwal, Slama, Ray, Schulman, Hilton, Kelton, Miller, Simens, Askell, Welinder, Christiano, Leike, and Lowe]{ouyang2022training}
Long Ouyang, Jeffrey Wu, Xu~Jiang, Diogo Almeida, Carroll~L. Wainwright, Pamela Mishkin, Chong Zhang, Sandhini Agarwal, Katarina Slama, Alex Ray, John Schulman, Jacob Hilton, Fraser Kelton, Luke Miller, Maddie Simens, Amanda Askell, Peter Welinder, Paul~F. Christiano, Jan Leike, and Ryan Lowe.
\newblock Training language models to follow instructions with human feedback.
\newblock In Sanmi Koyejo, S.~Mohamed, A.~Agarwal, Danielle Belgrave, K.~Cho, and A.~Oh (eds.), \emph{Advances in Neural Information Processing Systems 35: Annual Conference on Neural Information Processing Systems 2022, NeurIPS 2022, New Orleans, LA, USA, November 28 - December 9, 2022}, 2022.
\newblock URL \url{http://papers.nips.cc/paper\_files/paper/2022/hash/b1efde53be364a73914f58805a001731-Abstract-Conference.html}.

\bibitem[Park \& Kulkarni(2023)Park and Kulkarni]{park2024thinking}
Soya Park and Chinmay Kulkarni.
\newblock Thinking assistants: Llm-based conversational assistants that help users think by asking rather than answering.
\newblock \emph{CoRR}, abs/2312.06024, 2023.
\newblock \doi{10.48550/ARXIV.2312.06024}.
\newblock URL \url{https://doi.org/10.48550/arXiv.2312.06024}.

\bibitem[Pasquinelli et~al.(2021)Pasquinelli, Farina, Bedel, and Casati]{pasquinelli2021naturalizing}
Elena Pasquinelli, Mathieu Farina, Audrey Bedel, and Roberto Casati.
\newblock Naturalizing critical thinking: consequences for education, blueprint for future research in cognitive science.
\newblock \emph{Mind, Brain, and Education}, 15\penalty0 (2):\penalty0 168--176, 2021.

\bibitem[Pithers \& Soden(2000)Pithers and Soden]{pithers2000critical}
Robert~T Pithers and Rebecca Soden.
\newblock Critical thinking in education: A review.
\newblock \emph{Educational research}, 42\penalty0 (3):\penalty0 237--249, 2000.

\bibitem[Plato(369BCE)]{Plato2008Theaetetus}
Plato.
\newblock \emph{Theaetetus}.
\newblock 369BCE.

\bibitem[Plato(370BCE)]{Plato2008Phaedrus}
Plato.
\newblock \emph{Phaedrus}.
\newblock 370BCE.

\bibitem[Plato(380 BC)]{plato380bc_republic}
Plato.
\newblock \emph{The Republic}.
\newblock 380 BC.

\bibitem[Pock et~al.(2023)Pock, Ye, and Moore]{pock2023llmsgraspmoralityconcept}
Mark Pock, Andre Ye, and Jared Moore.
\newblock Llms grasp morality in concept, 2023.
\newblock URL \url{https://arxiv.org/abs/2311.02294}.

\bibitem[Popper(2002)]{Popper2002Logic}
Karl Popper.
\newblock \emph{The Logic of Scientific Discovery}.
\newblock Routledge Classics, 2nd edition, Feb 2002.
\newblock ISBN 978-0415278447.

\bibitem[Radensky et~al.(2024)Radensky, Weld, Chang, Siangliulue, and Bragg]{radensky2024letspointllmsupportedplanning}
Marissa Radensky, Daniel~S. Weld, Joseph~Chee Chang, Pao Siangliulue, and Jonathan Bragg.
\newblock Let's get to the point: Llm-supported planning, drafting, and revising of research-paper blog posts, 2024.
\newblock URL \url{https://arxiv.org/abs/2406.10370}.

\bibitem[Raths et~al.(1966)Raths, Wasserman, Jonas, and Rothstein]{raths1966teaching}
LE~Raths, S~Wasserman, A~Jonas, and A~Rothstein.
\newblock Teaching for critical thinking: Theory and application.
\newblock \emph{Columbus, OH: Charles-Merrill}, 1966.

\bibitem[Rawls(1971)]{41832fbd-4a82-3c18-ae99-e02c0f77c64f}
John Rawls.
\newblock \emph{A Theory of Justice: Original Edition}.
\newblock Harvard University Press, 1971.
\newblock ISBN 9780674880108.
\newblock URL \url{http://www.jstor.org/stable/j.ctvjf9z6v}.

\bibitem[Rayan et~al.(2024)Rayan, Kanetkar, Gong, Yang, Palani, Xia, and Dow]{Rayan2024}
Jude Rayan, Dhruv Kanetkar, Nicole Gong, Yuewen Yang, Srishti Palani, Haijun Xia, and Steven~P. Dow.
\newblock Exploring the potential for generative ai-based conversational cues for real-time collaborative ideation.
\newblock In \emph{Creativity and Cognition (C\&C ’24)}, pp.\ ~15, New York, NY, USA, June 23--26 2024. ACM.
\newblock \doi{10.1145/3635636.3656184}.

\bibitem[Reif(2008)]{reif2008applying}
Frederick Reif.
\newblock \emph{Applying cognitive science to education: Thinking and learning in scientific and other complex domains}.
\newblock MIT press, 2008.

\bibitem[Ricoeur(1981)]{ricoeur1981hermeneutics}
Paul Ricoeur.
\newblock \emph{Hermeneutics and the Human Sciences: Essays on Language, Action and Interpretation}.
\newblock Cambridge University Press, Cambridge, 1981.
\newblock ISBN 0521280028.

\bibitem[Rodman(2023)]{Rodman2023}
Emma Rodman.
\newblock On political theory and large language models.
\newblock \emph{Political Theory}, 2023.
\newblock \doi{10.1177/0090591723120082}.
\newblock URL \url{https://doi.org/10.1177/0090591723120082}.

\bibitem[Rozi{\`{e}}re et~al.(2023)Rozi{\`{e}}re, Gehring, Gloeckle, Sootla, Gat, Tan, Adi, Liu, Remez, Rapin, Kozhevnikov, Evtimov, Bitton, Bhatt, Canton{-}Ferrer, Grattafiori, Xiong, D{\'{e}}fossez, Copet, Azhar, Touvron, Martin, Usunier, Scialom, and Synnaeve]{rozière2024code}
Baptiste Rozi{\`{e}}re, Jonas Gehring, Fabian Gloeckle, Sten Sootla, Itai Gat, Xiaoqing~Ellen Tan, Yossi Adi, Jingyu Liu, Tal Remez, J{\'{e}}r{\'{e}}my Rapin, Artyom Kozhevnikov, Ivan Evtimov, Joanna Bitton, Manish Bhatt, Cristian Canton{-}Ferrer, Aaron Grattafiori, Wenhan Xiong, Alexandre D{\'{e}}fossez, Jade Copet, Faisal Azhar, Hugo Touvron, Louis Martin, Nicolas Usunier, Thomas Scialom, and Gabriel Synnaeve.
\newblock Code llama: Open foundation models for code.
\newblock \emph{CoRR}, abs/2308.12950, 2023.
\newblock \doi{10.48550/ARXIV.2308.12950}.
\newblock URL \url{https://doi.org/10.48550/arXiv.2308.12950}.

\bibitem[Sartre(1943)]{Sartre1993Being}
Jean-Paul Sartre.
\newblock \emph{Being and Nothingness: An Essay on Phenomenological Ontology}.
\newblock Éditions Gallimard, 1943.

\bibitem[Scherrer et~al.(2023)Scherrer, Shi, Feder, and Blei]{scherrer2023evaluating}
Nino Scherrer, Claudia Shi, Amir Feder, and David~M. Blei.
\newblock Evaluating the moral beliefs encoded in llms.
\newblock In Alice Oh, Tristan Naumann, Amir Globerson, Kate Saenko, Moritz Hardt, and Sergey Levine (eds.), \emph{Advances in Neural Information Processing Systems 36: Annual Conference on Neural Information Processing Systems 2023, NeurIPS 2023, New Orleans, LA, USA, December 10 - 16, 2023}, 2023.
\newblock URL \url{http://papers.nips.cc/paper\_files/paper/2023/hash/a2cf225ba392627529efef14dc857e22-Abstract-Conference.html}.

\bibitem[Schmitt \& Buschek(2021)Schmitt and Buschek]{Schmitt2021CharacterChatST}
Oliver Schmitt and Daniel Buschek.
\newblock Characterchat: Supporting the creation of fictional characters through conversation and progressive manifestation with a chatbot.
\newblock In Corina Sas, Neil A.~M. Maiden, Brian~P. Bailey, Celine Latulipe, and Ellen~Yi{-}Luen Do (eds.), \emph{Proceedings of the 13th {ACM} {SIGCHI} Conference on Creativity and Cognition, C{\&}C 2021, Virtual Event / Venice, Italy, June 22-23, 2021}, pp.\  7:1--7:10. {ACM}, 2021.
\newblock \doi{10.1145/3450741.3465253}.
\newblock URL \url{https://doi.org/10.1145/3450741.3465253}.

\bibitem[Schwitzgebel et~al.(2023)Schwitzgebel, Schwitzgebel, and Strasser]{schwitzgebel2023creating}
Eric Schwitzgebel, David Schwitzgebel, and Anna Strasser.
\newblock Creating a large language model of a philosopher.
\newblock \emph{CoRR}, abs/2302.01339, 2023.
\newblock \doi{10.48550/ARXIV.2302.01339}.
\newblock URL \url{https://doi.org/10.48550/arXiv.2302.01339}.

\bibitem[Schön(1987)]{Schon1987}
Donald~A. Schön.
\newblock \emph{Educating the Reflective Practitioner: Toward a New Design for Teaching and Learning in the Professions}.
\newblock Jossey-Bass, San Francisco, CA, 1987.
\newblock ISBN 978-1555420253.

\bibitem[Shaer et~al.(2024)Shaer, Cooper, Mokryn, Kun, and Ben~Shoshan]{shaer2024aiaugmentedbrainwritinginvestigatinguse}
Orit Shaer, Angelora Cooper, Osnat Mokryn, Andrew~L Kun, and Hagit Ben~Shoshan.
\newblock Ai-augmented brainwriting: Investigating the use of llms in group ideation.
\newblock In \emph{Proceedings of the CHI Conference on Human Factors in Computing Systems}, CHI '24, New York, NY, USA, 2024. Association for Computing Machinery.
\newblock ISBN 9798400703300.
\newblock \doi{10.1145/3613904.3642414}.
\newblock URL \url{https://doi.org/10.1145/3613904.3642414}.

\bibitem[Shin(2022)]{doi:10.1177/1461444821993801}
Donghee Shin.
\newblock The perception of humanness in conversational journalism: An algorithmic information-processing perspective.
\newblock \emph{New Media \& Society}, 24\penalty0 (12):\penalty0 2680--2704, 2022.
\newblock \doi{10.1177/1461444821993801}.
\newblock URL \url{https://doi.org/10.1177/1461444821993801}.

\bibitem[Shneiderman(2000)]{shneiderman2000creating}
Ben Shneiderman.
\newblock Creating creativity: user interfaces for supporting innovation.
\newblock \emph{ACM Transactions on Computer-Human Interaction (TOCHI)}, 7\penalty0 (1):\penalty0 114--138, 2000.

\bibitem[Shu et~al.(2023)Shu, Luo, Hoskere, Zhu, Liu, Tong, Chen, and Meng]{shu2023rewritelm}
Lei Shu, Liangchen Luo, Jayakumar Hoskere, Yun Zhu, Yinxiao Liu, Simon Tong, Jindong Chen, and Lei Meng.
\newblock Rewritelm: An instruction-tuned large language model for text rewriting, 2023.

\bibitem[Simon et~al.(2024)Simon, Rieder, and Branford]{simon2024philosophy}
J.~Simon, G.~Rieder, and J.~Branford.
\newblock The philosophy and ethics of ai: Conceptual, empirical, and technological investigations into values.
\newblock \emph{DISO}, 3\penalty0 (10), 2024.
\newblock \doi{10.1007/s44206-024-00094-2}.
\newblock URL \url{https://doi.org/10.1007/s44206-024-00094-2}.

\bibitem[Singer(1972)]{68204752-8786-318f-a155-4726d074c535}
Peter Singer.
\newblock Famine, affluence, and morality.
\newblock \emph{Philosophy and Public Affairs}, 1\penalty0 (3):\penalty0 229--243, 1972.
\newblock ISSN 00483915, 10884963.
\newblock URL \url{http://www.jstor.org/stable/2265052}.

\bibitem[Sorensen et~al.(2024)Sorensen, Moore, Fisher, Gordon, Mireshghallah, Rytting, Ye, Jiang, Lu, Dziri, Althoff, and Choi]{sorensen2024roadmap}
Taylor Sorensen, Jared Moore, Jillian Fisher, Mitchell~L. Gordon, Niloofar Mireshghallah, Christopher~Michael Rytting, Andre Ye, Liwei Jiang, Ximing Lu, Nouha Dziri, Tim Althoff, and Yejin Choi.
\newblock A roadmap to pluralistic alignment, 2024.
\newblock URL \url{https://doi.org/10.48550/arXiv.2402.05070}.

\bibitem[Spinoza(1677)]{Spinoza2003Ethics}
Baruch Spinoza.
\newblock \emph{Ethics, Demonstrated in Geometrical Order}.
\newblock Posthumous, 1677.

\bibitem[{Springer}(2024)]{springer_ai_editorial_policies}
{Springer}.
\newblock Artificial intelligence (ai) - editorial policies.
\newblock \url{https://www.springer.com/gp/editorial-policies/artificial-intelligence--ai-/25428500}, 2024.
\newblock Accessed: 2024-07-29.

\bibitem[Strachan et~al.(2024)Strachan, Albergo, Borghini, Pansardi, Scaliti, Gupta, Saxena, Rufo, Panzeri, Manzi, Graziano, and Becchio]{Strachan2024}
James W.~A. Strachan, Dalila Albergo, Giulia Borghini, Oriana Pansardi, Eugenio Scaliti, Saurabh Gupta, Krati Saxena, Alessandro Rufo, Stefano Panzeri, Guido Manzi, Michael S.~A. Graziano, and Cristina Becchio.
\newblock Testing theory of mind in large language models and humans.
\newblock \emph{Nature Human Behaviour}, 8\penalty0 (7):\penalty0 1285--1295, 2024.
\newblock \doi{10.1038/s41562-024-01882-z}.
\newblock URL \url{https://doi.org/10.1038/s41562-024-01882-z}.

\bibitem[Sun et~al.(2017)Sun, Yuan, Rosson, Wu, and Carroll]{10.1145/3027063.3053250}
Na~Sun, Chien Wen~(Tina) Yuan, Mary~Beth Rosson, Yu~Wu, and Jack~M. Carroll.
\newblock Critical thinking in collaboration: Talk less, perceive more.
\newblock In \emph{Proceedings of the 2017 CHI Conference Extended Abstracts on Human Factors in Computing Systems}, CHI EA '17, pp.\  2944–2950, New York, NY, USA, 2017. Association for Computing Machinery.
\newblock ISBN 9781450346566.
\newblock \doi{10.1145/3027063.3053250}.
\newblock URL \url{https://doi-org.offcampus.lib.washington.edu/10.1145/3027063.3053250}.

\bibitem[Tanprasert et~al.(2024)Tanprasert, Fels, Sinnamon, and Yoon]{10.1145/3613904.3642513}
Thitaree Tanprasert, Sidney~S Fels, Luanne Sinnamon, and Dongwook Yoon.
\newblock Debate chatbots to facilitate critical thinking on youtube: Social identity and conversational style make a difference.
\newblock In \emph{Proceedings of the CHI Conference on Human Factors in Computing Systems}, CHI '24, New York, NY, USA, 2024. Association for Computing Machinery.
\newblock ISBN 9798400703300.
\newblock \doi{10.1145/3613904.3642513}.
\newblock URL \url{https://doi.org/10.1145/3613904.3642513}.

\bibitem[Tanwar et~al.(2024)Tanwar, Shrivastva, Singh, and Kumar]{tanwar2024opinebot}
Henansh Tanwar, Kunal Shrivastva, Rahul Singh, and Dhruv Kumar.
\newblock Opinebot: Class feedback reimagined using a conversational {LLM}.
\newblock \emph{CoRR}, abs/2401.15589, 2024.
\newblock \doi{10.48550/ARXIV.2401.15589}.
\newblock URL \url{https://doi.org/10.48550/arXiv.2401.15589}.

\bibitem[Thomas(2006)]{Thomas2006}
David~R. Thomas.
\newblock A general inductive approach for analyzing qualitative evaluation data.
\newblock \emph{American Journal of Evaluation}, 27\penalty0 (2):\penalty0 237--246, 2006.
\newblock \doi{10.1177/1098214005283748}.

\bibitem[Thomson(2019)]{Thomson}
Iain~D. Thomson.
\newblock \emph{Rethinking the Analytic/Continental Divide}, pp.\  569–589.
\newblock Cambridge University Press, 2019.

\bibitem[Trouillot(1995)]{Winichakul1997MichelRolphTS}
Michel-Rolph Trouillot.
\newblock \emph{Silencing the Past: Power and the Production of History}.
\newblock Beacon Press books. Beacon Press, 1995.
\newblock ISBN 9780807043110.

\bibitem[Van~Dinter et~al.(2021)Van~Dinter, Tekinerdogan, and Catal]{van2021automation}
Raymon Van~Dinter, Bedir Tekinerdogan, and Cagatay Catal.
\newblock Automation of systematic literature reviews: A systematic literature review.
\newblock \emph{Information and Software Technology}, 136:\penalty0 106589, 2021.

\bibitem[Wagner et~al.(2022)Wagner, Lukyanenko, and Par{\'e}]{wagner2022artificial}
Gerit Wagner, Roman Lukyanenko, and Guy Par{\'e}.
\newblock Artificial intelligence and the conduct of literature reviews.
\newblock \emph{Journal of Information Technology}, 37\penalty0 (2):\penalty0 209--226, 2022.

\bibitem[Wang et~al.(2023{\natexlab{a}})Wang, Fu, Du, Gao, Huang, Liu, Chandak, Liu, Katwyk, Deac, Anandkumar, Bergen, Gomes, Ho, Kohli, Lasenby, Leskovec, Liu, Manrai, Marks, Ramsundar, Song, Sun, Tang, Veličković, Welling, Zhang, Coley, Bengio, and Zitnik]{Wang2023}
Hanchen Wang, Tianfan Fu, Yuanqi Du, Wenhao Gao, Kexin Huang, Ziming Liu, Payal Chandak, Shengchao Liu, Peter~Van Katwyk, Andreea Deac, Anima Anandkumar, Karianne Bergen, Carla~P. Gomes, Shirley Ho, Pushmeet Kohli, Joan Lasenby, Jure Leskovec, Tie-Yan Liu, Arjun Manrai, Debora Marks, Bharath Ramsundar, Le~Song, Jimeng Sun, Jian Tang, Petar Veličković, Max Welling, Linfeng Zhang, Connor~W. Coley, Yoshua Bengio, and Marinka Zitnik.
\newblock Scientific discovery in the age of artificial intelligence.
\newblock \emph{Nature}, 620\penalty0 (7972):\penalty0 47--60, August 1 2023{\natexlab{a}}.
\newblock \doi{10.1038/s41586-023-06221-2}.
\newblock URL \url{https://doi.org/10.1038/s41586-023-06221-2}.

\bibitem[Wang \& Tanes-Ehle(2022)Wang and Tanes-Ehle]{Wang2022}
Jinping Wang and Zeynep Tanes-Ehle.
\newblock Examining the effects of conversational chatbots on changing conspiracy beliefs about science: The paradox of interactivity.
\newblock \emph{Journal of Broadcasting \& Electronic Media}, 67\penalty0 (1):\penalty0 68--89, 2022.
\newblock \doi{10.1080/08838151.2022.2153842}.
\newblock URL \url{https://doi.org/10.1080/08838151.2022.2153842}.

\bibitem[Wang et~al.(2023{\natexlab{b}})Wang, Dong, Cheng, Liu, Yan, Gao, and Wei]{wang2023augmenting}
Weizhi Wang, Li~Dong, Hao Cheng, Xiaodong Liu, Xifeng Yan, Jianfeng Gao, and Furu Wei.
\newblock Augmenting language models with long-term memory.
\newblock In Alice Oh, Tristan Naumann, Amir Globerson, Kate Saenko, Moritz Hardt, and Sergey Levine (eds.), \emph{Advances in Neural Information Processing Systems 36: Annual Conference on Neural Information Processing Systems 2023, NeurIPS 2023, New Orleans, LA, USA, December 10 - 16, 2023}, 2023{\natexlab{b}}.
\newblock URL \url{http://papers.nips.cc/paper\_files/paper/2023/hash/ebd82705f44793b6f9ade5a669d0f0bf-Abstract-Conference.html}.

\bibitem[White(1973)]{White1975MetahistoryTH}
Hayden White.
\newblock \emph{Metahistory: The Historical Imagination in Nineteenth-century Europe}.
\newblock Johns Hopkins University, 1973.

\bibitem[Xu et~al.(2024)Xu, Yin, Gu, Mar, Zhang, E, and Dow]{10.1145/3640543.3645196}
Xiaotong~(Tone) Xu, Jiayu Yin, Catherine Gu, Jenny Mar, Sydney Zhang, Jane~L. E, and Steven~P. Dow.
\newblock Jamplate: Exploring llm-enhanced templates for idea reflection.
\newblock In \emph{Proceedings of the 29th International Conference on Intelligent User Interfaces}, IUI '24, pp.\  907–921, New York, NY, USA, 2024. Association for Computing Machinery.
\newblock ISBN 9798400705083.
\newblock \doi{10.1145/3640543.3645196}.
\newblock URL \url{https://doi.org/10.1145/3640543.3645196}.

\bibitem[Yamamoto \& Yamamoto(2018)Yamamoto and Yamamoto]{10.1145/3176349.3176377}
Yusuke Yamamoto and Takehiro Yamamoto.
\newblock Query priming for promoting critical thinking in web search.
\newblock In \emph{Proceedings of the 2018 Conference on Human Information Interaction \& Retrieval}, CHIIR '18, pp.\  12–21, New York, NY, USA, 2018. Association for Computing Machinery.
\newblock ISBN 9781450349253.
\newblock \doi{10.1145/3176349.3176377}.
\newblock URL \url{https://doi.org/10.1145/3176349.3176377}.

\bibitem[Yuan et~al.(2022)Yuan, Coenen, Reif, and Ippolito]{10.1145/3490099.3511105}
Ann Yuan, Andy Coenen, Emily Reif, and Daphne Ippolito.
\newblock Wordcraft: Story writing with large language models.
\newblock In \emph{Proceedings of the 27th International Conference on Intelligent User Interfaces}, IUI '22, pp.\  841–852, New York, NY, USA, 2022. Association for Computing Machinery.
\newblock ISBN 9781450391443.
\newblock \doi{10.1145/3490099.3511105}.
\newblock URL \url{https://doi.org/10.1145/3490099.3511105}.

\bibitem[Zarouali et~al.(2021)Zarouali, Makhortykh, Bastian, and Araujo]{doi:10.1177/0267323120940908}
Brahim Zarouali, Mykola Makhortykh, Mariella Bastian, and Theo Araujo.
\newblock Overcoming polarization with chatbot news? investigating the impact of news content containing opposing views on agreement and credibility.
\newblock \emph{European Journal of Communication}, 36\penalty0 (1):\penalty0 53--68, 2021.
\newblock \doi{10.1177/0267323120940908}.
\newblock URL \url{https://doi.org/10.1177/0267323120940908}.

\bibitem[Zhao(2022)]{Zhao2022LeveragingAI}
Xin Zhao.
\newblock Leveraging artificial intelligence (ai) technology for english writing: Introducing wordtune as a digital writing assistant for efl writers.
\newblock \emph{RELC Journal}, 54:\penalty0 890 -- 894, 2022.

\bibitem[Zhao et~al.(2024)Zhao, Yan, Sun, Xing, Wang, Meng, Cheng, Ren, and Yin]{zhao2024improving}
Yukun Zhao, Lingyong Yan, Weiwei Sun, Guoliang Xing, Shuaiqiang Wang, Chong Meng, Zhicong Cheng, Zhaochun Ren, and Dawei Yin.
\newblock Improving the robustness of large language models via consistency alignment.
\newblock In Nicoletta Calzolari, Min{-}Yen Kan, V{\'{e}}ronique Hoste, Alessandro Lenci, Sakriani Sakti, and Nianwen Xue (eds.), \emph{Proceedings of the 2024 Joint International Conference on Computational Linguistics, Language Resources and Evaluation, {LREC/COLING} 2024, 20-25 May, 2024, Torino, Italy}, pp.\  8931--8941. {ELRA} and {ICCL}, 2024.
\newblock URL \url{https://aclanthology.org/2024.lrec-main.782}.

\bibitem[Ziems et~al.(2023)Ziems, Dwivedi{-}Yu, Wang, Halevy, and Yang]{Ziems2023NormBankAK}
Caleb Ziems, Jane Dwivedi{-}Yu, Yi{-}Chia Wang, Alon~Y. Halevy, and Diyi Yang.
\newblock Normbank: {A} knowledge bank of situational social norms.
\newblock In Anna Rogers, Jordan~L. Boyd{-}Graber, and Naoaki Okazaki (eds.), \emph{Proceedings of the 61st Annual Meeting of the Association for Computational Linguistics (Volume 1: Long Papers), {ACL} 2023, Toronto, Canada, July 9-14, 2023}, pp.\  7756--7776. Association for Computational Linguistics, 2023.
\newblock \doi{10.18653/V1/2023.ACL-LONG.429}.
\newblock URL \url{https://doi.org/10.18653/v1/2023.acl-long.429}.

\end{thebibliography}
\bibliographystyle{colm2024_conference}

\newpage 

\appendix
\section{Interviewee Information Sheet}
\label{info-sheet}

Table~\ref{tab:participant_info} provides high-level information about each interviewee which may be relevant to interpreting and contextualizing their views.
The \textit{General Interest(s)} feature describes the broad fields that the interviewees work in.
The \textit{Notable Specific Interest(s)} feature describes any specific topics in the field(s) mentioned in the \textit{General Interest(s)} feature that the interviewees focus their work on. This feature is not exclusive, meaning that interviewees may also work on other topics outside of the specific interests. If the value for this feature is blank, then the interviewee's work is sufficiently characterized by the value in the \textit{General Interest(s)} feature.
The \textit{Experience with LMs} feature describes three levels of experience with using LMs: little to none, limited, and extensive.
If interviewees have either limited or extensive experience with using LMs, the \textit{Uses of LMs} feature describes their primary use: for teaching (e.g., using LMs to teach material, trying to understand features of LM-generated student submissions), for personal use (e.g., to improve productivity, for entertainment), for exploration (i.e., playing around with the LM out of curiosity to understand the technology better), and for research (i.e., their research is on LMs).
Note that the following interviewees have published at least one article on some aspect of AI: \icite{P5, P6, P13, P14}.

\begin{sidewaystable}[!htbp]
\centering
\small
\begin{tabular}{@{}p{0.05\textwidth}p{0.15\textwidth}p{0.2\textwidth}p{0.2\textwidth}p{0.15\textwidth}p{0.15\textwidth}@{}}
\toprule
ID & Title & General Interest(s) & Notable Specific Interest(s) & Experience with LMs & Use of LMs \\
\midrule
P1 & Associate Professor & Ethics, Political Philosophy & Bioethics, Feminist Ethics & Limited & For teaching \\
P2 & Associate Professor & Philosophy of Science & Philosophy of Biology & Limited & For exploration \\
P3 & Professor & Ethics, Aesthetics & Meta-ethics & Limited & For teaching \\
P4 & Professor & Ethics, Political Philosophy & & Limited & For personal use \\
P5 & Assistant Professor & Ethics & Virtue ethics & Limited & For teaching \\
P6 & Assistant Professor & Ethics, Political Philosophy & Philosophy of Technology, AI & Extensive & For research \\
P7 & Assistant Professor & Philosophy of Science & Philosophy of Physics & Extensive & For personal use \\
P8 & Associate Professor & History of Philosophy & German philosophy & Limited & For personal use \\
P9 & Professor & Philosophy of Science & & Extensive & For exploration \\
P10 & Associate Professor & Ethics, History of Philosophy & Philosophy of Technology & Little to None & \\
P11 & Professor & Philosophy of Science & Philosophy of Statistics & Little to None & \\
P12 & Professor & Philosophy of Science & Psychology & Limited & For exploration \\
P13 & Associate Professor & Philosophy of Science & Philosophy of Biology & Limited & For class \\
P14 & Professor & Logic, Philosophy of Mind & Semantics, Linguistics & Extensive & For exploration \\
P15 & Assistant Professor & Aesthetics & Value theory, Literature & Limited & For exploration \\
P16 & Professor & Ethics, Political Philosophy & Public and Global Policy & Limited & For exploration \\
P17 & Teaching Professor & Pedagogy, Epistemology & & Extensive & For exploration \\
P18 & Associate Professor & Philosophy of Science & Philosophy of Physics & Limited & For class \\
P19 & Assistant Professor & Ethics & Moral psychology & Limited & For personal use \\
P20 & Professor & History of Philosophy & & Little to None & \\
P21 & Associate Professor & Ethics & Bioethics & Little to None & \\
\bottomrule
\end{tabular}
\caption{Interviewee information.}
\label{tab:participant_info}
\end{sidewaystable}

\newpage

\section{Interview Questions and Guidelines}
\label{interview}


\begin{enumerate}
    \item Meta-philosophy
    \begin{enumerate}
        \item What is philosophy? Why do you go about doing philosophy? What aims do you have?
        \item What drives the ‘doing’ of philosophy? What is the role of personal motivations, subjective experience, and aesthetic judgements?
        \item Who or what can ‘do’ philosophy? For instance, can LLMs ‘do’ philosophy?
        \item What makes doing philosophy ‘difficult’ / nontrivial?
        \item How does philosophy distinguish its products from those of other disciplines?
    \end{enumerate}
    \item The philosophical process
    \begin{enumerate}
        \item How do you go from no idea to a spark of an idea / an unrefined idea?
        \item How do you develop and refine philosophical ideas? What moves have to happen?
        \item How mechanical / creative is the process of doing philosophy?
        \item What is the relationship between texts / textual methods and philosophy? Does philosophizing, to some extent, operate ‘above’ language in ideas / thoughts?
        \item What is the role of conversation in the doing of philosophy? What are some of its challenges?
        \item What makes for a good interlocutor, and what makes for a good conversation?
    \end{enumerate}
    \item Language Models for philosophy
    \begin{enumerate}
        \item What roles can language models play in the development of philosophy?
        \item What do language models need to be better in the development of philosophy?
        \item What are some of the opportunities and strengths for language models in philosophy?
        \item What are some of the risks and weaknesses for language models in philosophy?
        \item Would you use language models in intellectually substantive ways currently? What about in the future, with plausible improvements?
    \end{enumerate}
\end{enumerate}

\end{document}